\documentclass[%
 reprint,
  aps,
 prl,
]{revtex4-2}

\usepackage{graphicx}
\usepackage{xcolor}%
\usepackage{dcolumn}
\usepackage{bm}

\def\ZZ{\mathrm{Z}}
\def\AC{\mathrm{A}}
\def\Step{\mathrm{Edge}}
\def\thetaZZ{\theta_\ZZ}

\def\thetaStep{\theta_\Step}
\def\GAC{G_\AC}
\def\GZZ{G_\ZZ}
\def\gammaAC{\gamma_\AC}
\def\gammaZZ{\gamma_\ZZ}
\def\Kone{K_\mathrm{\uppercase\expandafter{\romannumeral1}}}
\def\Ktwo{K_\mathrm{\uppercase\expandafter{\romannumeral2}}}
\def\Gc{G_\mathrm{c}}

\begin{document}


\title{Non-Equilibrium Nature of Fracture Determines the Crack Paths}

\author{Pengjie Shi}
\author{Shizhe Feng}%
\author{Zhiping Xu}%
\email{xuzp@tsinghua.edu.cn}
\affiliation{%
 Applied Mechanics Laboratory and Department of Engineering Mechanics, Tsinghua University, Beijing, 100084, China.
}%

\date{\today}

\begin{abstract}
A high-fidelity neural network-based force field, NN-F$^{3}$, is developed to cover the strain states up to material failure and the non-equilibrium, intermediate nature of fracture.
Simulations of fracture in 2D crystals using NN-F$^{3}$ reveal spatial complexities from lattice-scale kinks to sample-scale patterns.
We find that the fracture resistance cannot be quantified by the energy densities of relaxed edges as in the literature.
Instead, the fracture patterns, critical stress intensity factors at the kinks, and energy densities of edges in the intermediate, unrelaxed states offer reasonable measures for the fracture toughness and its anisotropy.
\end{abstract}

\maketitle

\begin{figure}[bht]
{
\includegraphics[width=8.6cm]{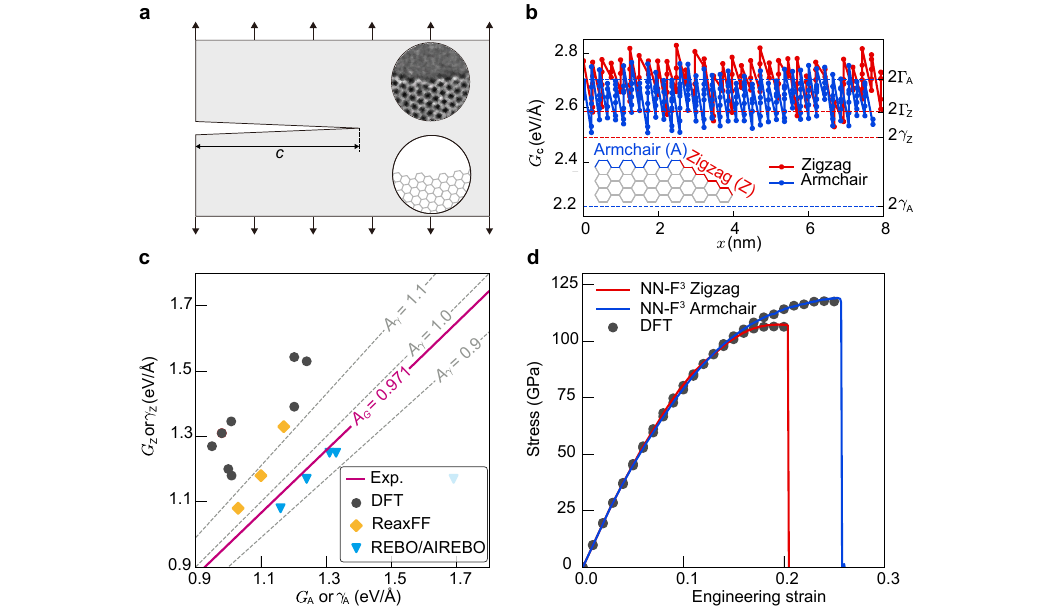}
\caption{
(a) A crack-containing graphene specimen under force loading~\cite{girit_2009}.
(b) Fracture toughness $\Gc$ of cracks along the zigzag and armchair motifs, $\Gc=\Kone^2/E$~\cite{supp_info}.
The oscillation in $\Gc$ along with the advancing cracks signals the lattice trapping events.
(c) Theoretical predictions of $A_\gamma$ \cite{Boris_2010, zhang_2014, yin_2015, jun_2008, koskinen_2008, huang_2009, gan_2010, gao_2011, sen_2010, kim_2012, song_2017}, and experimental measurements of $A_G$ \cite{feng_2022,qu_2022}.
(d), Uniaxial stress-strain relations measured along the zigzag and armchair directions.
}\label{Fig_1}
}
\end{figure}

Fracture is a catastrophic process in nature and engineering, which leaves facets and kinks along the crack paths. 
In 2D crystals such as graphene and h-BN, different types of edges can be cleaved by fracture, including zigzag (Z), armchair (A), and mixed zigzag-armchair or chiral (C) edges (Fig.\,\ref{Fig_1}a).
The relative stabilities of graphene edge structures were explored experimentally through the abundance of edges created by various techniques such as fracture \cite{feng_2022}, mechanical exfoliation \cite{qu_2022}, and irradiation \cite{fujihara_2015}. 
Most observations show almost the same probabilities of zigzag and armchair edges \cite{kim_2013,neubeck_2010,jia_2009}, while a few of them report either zigzag or armchair direction is preferred over the other \cite{fujihara_2015,shi_2020,girit_2009,qu_2022,kim_2012}.
These facts suggest that the stabilities of zigzag and armchair edges could be quite close, which contradicts theoretical predictions from first-principles calculations.
Ground-state calculations based on the density functional theory (DFT) show that electronic and structural relaxation of the armchair edge of graphene significantly reduces its energy density, $\gammaAC$, which becomes lower than that of the zigzag edge, $\gammaZZ$ \cite{jun_2008,koskinen_2008,huang_2009,gan_2010,Boris_2010,gao_2011,yin_2015}.
The disagreement between theory and experiment remains unsolved for more than a decade.

The selection of crack paths during fracture is closely related to the relative stability of edges. 
Under the framework of fracture mechanics, the crack driving force can be measured by the energy release rate (ERR), $G$, while the energy cost to activate the fracture is defined as the fracture toughness, $\Gc$ \cite{lawn_2004}.
The value of $\Gc$ is difficult to determine by theory due to the non-equilibrium nature of fracture and thus commonly measured by experiments \cite{lawn_2004}.
In theoretical studies, the fracture resistance is usually approximated by the surface or edge energy densities as $\Gc = 2\gamma$ (Fig.\,\ref{Fig_1}b) \cite{griffith_1921,lawn_2004,zhigong_2021,zhang_2014}.
The directional dependence of $\Gc\left(\theta\right)$, which defines the relative stability of different edges, is expected to align with that of $2\gamma\left(\theta\right)$ \cite{takei_2013,feng_2022}.

By analyzing the crack path under specific loading conditions, the relative stability or the anisotropy of fracture can be deduced.
The anisotropy in $\Gc\left(\theta\right)$ and $\gamma\left(\theta\right)$ of crystals with a honeycomb lattice such as graphene can be quantified by the ratios between their values at post-fracture zigzag (Z) and armchair (A) edges, $A_G=\GZZ/\GAC$, and $A_\gamma=\gammaZZ/\gammaAC$ \cite{Boris_2010}, respectively \cite{kim_2013, neubeck_2010, jia_2009}.
Recently, direct tensile tests of monolayer graphene and peeling tests of highly-oriented pyrolytic graphite (HOPG), 
although unable to resolve the atomic-scale edge structures,
conclude weak anisotropies ($A_G = 1.06$ \cite{feng_2022} and $0.971$ \cite{qu_2022}) from the overall orientation of cracks.

Energy densities of edges cleaved by fracture cannot be directly measured in experiments, and the use of theoretically calculated $2\gamma(\theta)$ as the fracture toughness remains questionable.
In fact, experimentally measured fracture toughness is usually much higher than the value of $2\gamma$ even for brittle crystals where plastic dissipation is absent \cite{zhigong_2021,zhang_2014,delrio_2015, delrio_2022} (Fig.\,\ref{Fig_1}b).
Large-scale molecular dynamics (MD) simulations may help address the issue if provided with force fields of high accuracy and low cost.
Empirical force fields (FFs) reported in the literature cannot capture the non-equilibrium nature and high, non-uniform lattice distortion at the crack front \cite{jung_2019,Boris_2010, sen_2010,kim_2012, zhang_2014, yin_2015, song_2017, zhang2022atomistic}. 
The values of $A_\gamma$ calculated using Stillinger-Weber or Tersoff potentials are the same as the bond-cutting estimation, $\sqrt{3}/2$, and do not correctly capture the bonding characteristics of materials \cite{hossain_2018,zhang2022atomistic}.
By including the chemistry of interatomic bonding, the adaptive intermolecular reactive empirical bond order (AIREBO) predicts $A_\gamma < 1$ \cite{Boris_2010, zhang_2014, yin_2015}, while the reactive force field (ReaxFF) yields opposite results, $A_\gamma>1$  \cite{sen_2010,kim_2012,song_2017} (Fig.\,\ref{Fig_1}c).
Compared with the experimental measurements \cite{feng_2022,qu_2022}, DFT calculations predict a relatively strong anisotropy with $A_\gamma>1.1$, where electronic and structural relaxation of the edges are taken into account \cite{jun_2008,koskinen_2008,huang_2009,gan_2010,Boris_2010,gao_2011,yin_2015}\,\cite{supp_info}.
The DFT predictions are also quantitatively different from the AIREBO and ReaxFF results.
Recently, the implementation of neural network-based force fields \cite{friederich_2021} shows the capability to resolve the accuracy-cost dilemma and led to significant progress in several fields \cite{galib2021reactive, font2022predicting, li2022origin, hedman2023dynamics}.
However, the lack of a reasonable description of the non-equilibrium nature and exploration of the full space of strain states leaves the atomistic approach to fracture still immature.

Here we develop a neural network-based force field for fracture (NN-F$^{3}$) for 2D crystals including graphene and h-BN based on first-principles calculations and an active-learning framework \cite{friederich_2021}.
The tensorial nature of strain states and the undercoordination nature of cleavaged edges~\cite{Boris_2010} demand a large training set of DFT data and are addressed by an active-learning workflow~\cite{dpgen}.
Our training sets for NN-F$^{3}$ include structures with strained lattices (the uniaxial strain in the range of $0-0.25$ along different lattice directions), cleaved edges (zigzag and armchair segments as well as the kinks between them), and cracks ($203,554$ datasets in total).
The Deep Potential Smooth Edition (DeepPot-SE) model~\cite{NEURIPS2018_e2ad76f2,deepmd} is used to train NN-F$^{3}$, 
the performance of which is validated by reporting mean absolute errors (MAEs) of the energies per atom, the edge energy densities, and the interatomic forces below $2\,\mathrm{meV}$, $2.2\,\mathrm{meV/\AA}$ and $43\,\mathrm{meV/\AA}$, respectively.
The relative error (RE) in the stress-strain relations is under $2\%$ (Fig.\,\ref{Fig_1}d).
The workflow thus assures an accurate description of the crack-tip processes\,\cite{supp_info}.

\begin{figure}[bht]%
\centering
\includegraphics[width=8.6cm]{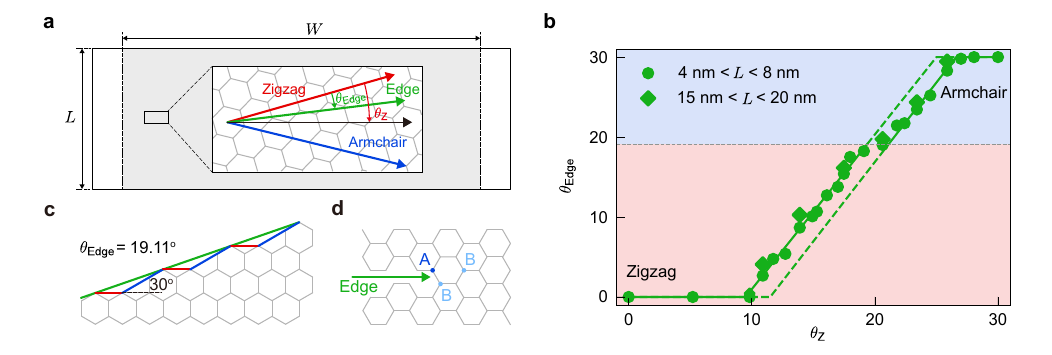}
\caption{
(a) The simulation setup of fracture tests.
(b) The relation between $\thetaZZ$ and $\thetaStep$ obtained from the MD simulations for graphene using NN-F$^{3}$ and theoretical predictions using Eq.\,\ref{Eq_2}.
(c) The atomic-level structure of edges with $\thetaStep=19.11^\circ$.
(d) The asymmetry between A- and B-site atoms along armchair edges at the crack tip.
}
\label{Fig_2}
\end{figure}

\textit{Fracture Anisotropy of Graphene}-
The fracture of graphene is explored by quasi-static uniaxial tension using the Large-scale Atomic/Molecular Massively Parallel Simulator (LAMMPS) \cite{lammps,supp_info}.
In order to host relatively long cracks, wide samples ($W\approx50\,\mathrm{nm}$) are constructed (Fig.\,\ref{Fig_2}).
One atom at the left edge is removed to initialize the crack.
Periodic boundary conditions (PBCs) are enforced along the tensile direction.
The span $L$ is in the range of $4 - 8~\mathrm{nm}$ (and $15 - 20~\mathrm{nm}$ to see the size dependence) to accommodate different lattice orientations ($\thetaZZ\in\left[0^\circ,30^\circ\right]$, measured from the zigzag motif, Fig.\,\ref{Fig_2}a, b).

\begin{figure*}[bht]%
\centering
\includegraphics[width=17.8cm]{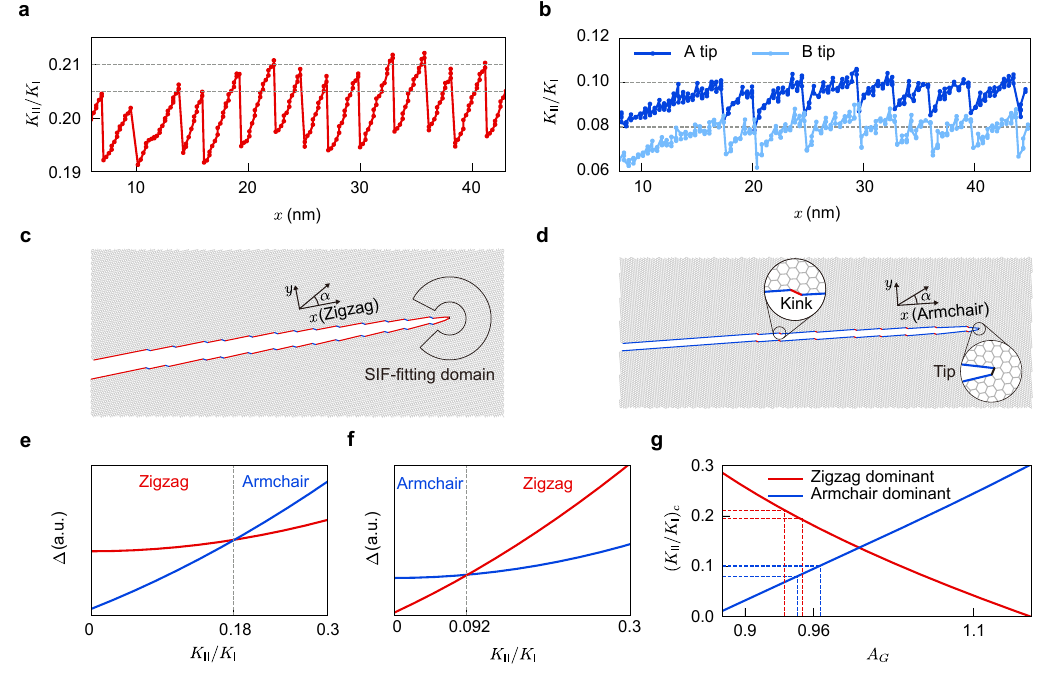}
\caption{
The ratio of SIFs ($\Ktwo/\Kone$) along with the advancing crack at $\thetaZZ=10.89^\circ$ (a) and $\thetaZZ=25.87^\circ$ (b).
(c and d) Cleaved edges with zigzag and armchair segments.
The domain defined to calculate the SIFs is annotated~\cite{supp_info}.
(e and f) Theoretical predictions of the relations between $\Delta$ and $\Ktwo/\Kone$ from Eq.\,\ref{Eq_4} using $A_G=0.96$.
(g) Theoretical predictions of the relation between $A_G$ and $\left(\Ktwo/\Kone\right)_c$ using Eq.\,\ref{Eq_4}.
}
\label{Fig_3}
\end{figure*}

MD simulation results show that the cracks are straight at the sample scale (Figs.\,\ref{Fig_3})~\cite{supp_info}.
Their overall orientations are denoted by $\thetaStep$ (measured from the zigzag motif, see Fig.\,\ref{Fig_2}a).
However, the cracks may deflect at the lattice scale, leaving kinks between the zigzag and armchair segments behind (Fig.\,\ref{Fig_2}c).
The relations between $\thetaStep$ and $\thetaZZ$ are summarized in Fig.\,\ref{Fig_2}b, which can be classified into three regimes.
For $\thetaZZ\in[0^\circ,10^\circ]$ or $[27^\circ,30^\circ]$, the crack advances along the zigzag ($\thetaStep=0^\circ$) or armchair ($\thetaStep=30^\circ$) direction, respectively.
For $\thetaZZ\in[10^\circ,27^\circ]$, the crack advances between them ($\thetaStep\in\left(0^\circ,30^\circ\right)$).
Cleavage of zigzag edges dominates if the loading direction is uniformly sampled, which is attributed to the fact of $\GZZ<\GAC$\,\cite{supp_info}.
This finding conforms with the observations in the peeling experiments of HOPG where the polycrystalline texture is randomly oriented \cite{qu_2022}.
However, the energy densities of relaxed edge $\gamma(\theta)$ obtained from DFT calculations display an opposite trend of $\gammaZZ > \gammaAC$ \cite{jun_2008, koskinen_2008, huang_2009, gan_2010, Boris_2010, gao_2011, yin_2015, supp_info}.
This inconsistency indicates that $\gamma(\theta)$ fails to correctly characterize the anisotropy in fracture resistance.

The crack driving force under uniaxial tensile stress $\sigma_y$ is $G\left(\thetaStep\right)\sim\cos^2\left(\thetaZZ-\thetaStep\right)\sigma_y^2$\,\cite{supp_info}.
Following the criterion of maximum ERR (MERR), the crack will advance in the direction with $G\left(\thetaStep\right) \geq \Gc\left(\thetaStep\right)$. 
In the honeycomb lattice of graphene, the cleaved edges consist of zigzag and armchair segments, and the value of $\Gc(\thetaStep)$ can be estimated as the average value of $\GAC$ and $\GZZ$ weighted by their lengths \cite{qu_2022,Boris_2010}, that is
\begin{equation}
\Gc\left(\thetaStep\right)=2\GAC\left[\sin\left(\thetaStep\right)+A_G\sin\left(30^\circ-\thetaStep\right)\right].\label{Eq_1}
\end{equation}
This result presumes that the formation and interaction energies of lattice kinks are negligible in comparison with the edge energies \cite{Boris_2010, lee2023importance}.
The direction of crack propagation, $\thetaStep$, can thus be obtained from the lattice orientation, $\thetaZZ$, by finding the minimum of $\sigma_y^2$ that satisfies $G = \Gc$, that is
\begin{equation}
	\thetaStep = \arg\min\sigma_y^2 = \arg\min\frac{\Gc\left(\thetaStep\right)}{\cos^2\left(\thetaZZ-\thetaStep\right)}.\label{Eq_2}
\end{equation}
The predictions using $A_G=0.96$ and $0.93$ fit the simulation results for $\thetaStep$ smaller and larger than the critical value of $\theta_{\rm c} = 19.11^\circ$ (Fig.\,\ref{Fig_2}b), respectively.
At $\thetaStep=19.11^\circ$, the numbers of zigzag and armchair segments are the same along the edge (Fig.\,\ref{Fig_2}c).
The smaller values of $A_G$ at $\thetaStep>19.11^\circ$ may be attributed to the asymmetry between the A and B sites at the armchair edges (Fig.\,\ref{Fig_2}d), which can elevate the fracture toughness and promote deflection.
Similar effects of the edge asymmetry on crack deflection and toughening were also observed in h-BN \cite{zhigong_2021,supp_info} and $\mathrm{WS_2}$ \cite{jung_2019}.

\textit{Origin of Edge Kinks}-
The high-fidelity NN-F$^{3}$ allows us to explore the edge structures at the atomic level.
Zigzag and armchair segments as well as lattice kinks connecting them can be resolved at length scales where fracture mechanics can be applied for analysis.
Large-scale MD simulations excluding the size effects show periodic crack patterns\,\cite{supp_info}
, and highlight the advantages of NN-F$^{3}$ in offering high accuracy at the first-principles level and low computation cost that allows direct simulations up to the experimental scale\,\cite{supp_info,feng_2022} 
The criterion of MERR \cite{nuismer1975,takei_2013,feng_2022} suggests that the direction of a propagating crack defined in the local coordinate system (Fig.\,\ref{Fig_3}c and \ref{Fig_3}d) is

\begin{equation}
	\alpha = \arg\max\frac{G\left(\alpha\right)}{\Gc\left(\alpha\right)},\label{Eq_3}
\end{equation}
where $G(\alpha)$ is evaluated by the SIFs in the tensile and shear modes ($\Kone$ and $\Ktwo$, respectively) as the out-of-plane displacement is ignored \cite{song_2017}.
The values of $\Kone$ and $\Ktwo$ can be determined by fitting the crack-tip displacement field with the Williams power expansion \cite{williams1957,supp_info}.

For loading conditions with crack directions not aligning with the zigzag or armchair motifs, we find that the presence of a mode-{\uppercase\expandafter{\romannumeral2}} feature could deflect the mode-{\uppercase\expandafter{\romannumeral1}} crack \cite{lawn_2004, feng_2022}.
The effect can be measured by the ratio $\Ktwo/\Kone$ extracted from MD simulations.
Two representative examples are shown in Figs.\,\ref{Fig_3}c and \ref{Fig_3}d, where the cleaved edges are dominated by the zigzag and armchair segments, respectively.
The value of $\Ktwo/\Kone$ oscillates as the crack propagates (Figs.\,\ref{Fig_3}a and \ref{Fig_3}b), indicating that the deflection is activated as the ratio approaches the threshold values $\left(\Ktwo/\Kone\right)_\mathrm{c}$.
The threshold depends on the loading conditions and lattice orientations and is higher for cracks advancing in the zigzag direction than that along the armchair ones.
The asymmetry between the A and B sites of the armchair edge further breaks the symmetry (Fig.\,\ref{Fig_2}d) and yields two thresholds (Fig.\,\ref{Fig_3}b), which confirms the effect of edge asymmetry on $A_{G}$ (Fig.\,\ref{Fig_2}b). 

To estimate the values of $\left(\Ktwo/\Kone\right)_\mathrm{c}$ and their relations with $A_G$, a dimensionless quantity $\Delta$ is introduced based on Eq.\,\ref{Eq_3} as\,\cite{supp_info}
\begin{equation}
	\alpha = \arg\max\left[\frac{G\left(\alpha\right)}{\Gc\left(\alpha\right)}\frac{2\GAC E}{\Kone^2}\right]= \arg\max\left[\Delta\left(\alpha,A_G,\frac{\Ktwo}{\Kone}\right)\right],\label{Eq_4}
\end{equation}
The direction of cracks determined from $A_G$ and $\Ktwo/\Kone$ follows the armchair or zigzag motifs ($\alpha = 0^\circ$ or $30^\circ$) due to the discrete nature of lattices. 
The relations between $\Ktwo/\Kone$ and $\Delta$ in the armchair- and zigzag-dominated regimes with $A_G=0.96$ are summarized in Figs.\,\ref{Fig_3}e and \ref{Fig_3}f, where the thresholds $\left(\Ktwo/\Kone\right)_\mathrm{c}$ are identified as $0.092$ and $0.18$, respectively.
Alternatively, the values of $A_G$ can be obtained from $\left(\Ktwo/\Kone\right)_\mathrm{c}$ that is directly determined by experiments or simulations (Fig.\,\ref{Fig_3}g and Table\,\ref{Tab_1}).
The results show that the value of $A_G$ does not match the anisotropy measured from the energies of relaxed edges in direct NN-F$^{3}$ or DFT calculations, $A_\gamma=\gammaZZ/\gammaAC = 1.113$ (Figs.\,\ref{Fig_1}c)~\cite{supp_info}.

\begin{figure}[bht]%
\centering
\includegraphics[width=8.6cm]{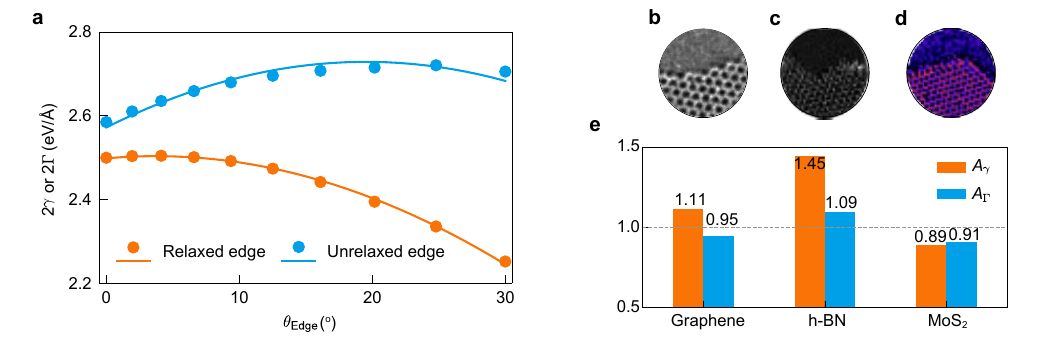}
\caption{
(a) Energy densities of relaxed ($\gamma$) and unrelaxed ($\Gamma$) graphene edges.
TEM images of edges in 2D graphene (b), h-BN (c), and MoS$_2$ (d) crystals adapted from \cite{girit_2009,zhigong_2021,wang_2016}.
(e) Values of $A_\gamma$ and $A_\Gamma$ of graphene, h-BN and MoS$_2$ obtained from DFT calculations.}
\label{Fig_4}
\end{figure}

\begin{table}[b]
\caption{The number of kinks in the sample with width $W \approx 50$ nm, and the relation between $\left(\Ktwo/\Kone\right)_{\rm c}$ and $A_G$ under loading conditions defined by $\thetaZZ$.} \label{Tab_1}%
\begin{ruledtabular}
\begin{tabular}{cccc}
\textrm{$\thetaZZ$}&
\textrm{\# kinks }&
\textrm{$\left(\Ktwo/\Kone\right)_{\rm c}$}&
\textrm{$A_G$}\\
\colrule
$25.87^\circ$ & 8  & $\left[0.080,0.100\right]$  & $\left[0.940,0.966\right]$  \\
$10.89^\circ$ & 13 &$\left[0.205,0.210\right]$   & $\left[0.935,0.940\right]$   \\
$9.82^\circ$  &  2 & $\left[0.195,0.200\right]$ & $\left[0.945,0.950\right]$  \\  
\end{tabular}
\end{ruledtabular}
\end{table}

\begin{figure}[bht]%
\centering
\includegraphics[width=8.7cm]{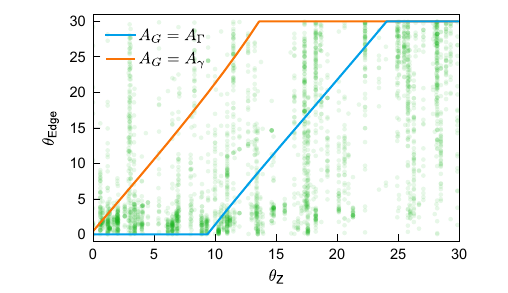}
\caption{
$\thetaStep-\thetaZZ$ relations of graphene obtained by assuming $A_G=A_{\gamma}$ (dash line) and $A_G=A_{\Gamma}$ (solid line)~\cite{supp_info}.
The data points are measurements from the peeling experiments of HOPG \cite{qu_2022}.
}
\label{Fig_5}
\end{figure}

\textit{Energy Densities of Unrelaxed Edges}-
The mismatch between $A_{G}$ and $A_{\gamma}$ implies that the anisotropy in the edge energies density fails to capture the atomistic kinetics of fracture, which selects the crack path.
Since the work of fracture should not depend on posterior edge relaxation processes after the event of cleavage, the energy densities of unrelaxed edges, $2\Gamma(\theta)$, are calculated using NN-F$^{3}$ or DFT calculations and compared to $2\gamma(\theta)$ for relaxed edges.
The results, $A_{\Gamma}=0.959<1$, suggest a weak anisotropy in the fracture toughness, agreeing well with the experimental evidence \cite{qu_2022,feng_2022} and the simulation results (Fig.\,\ref{Fig_4}a).
The values of $2\Gamma_{\rm Z}$ and $2\Gamma_{\rm A}$ also conform qualitatively well with $G_{\rm Z}$ and $G_{\rm A}$, respectively, by ignoring the lattice-trapping effects (Fig.\,\ref{Fig_1}b).
We investigate several measures of fracture energies\,\cite{supp_info}, 
and conclude that the energy density of unrelaxed edges, $A_\Gamma$, can be a good indicator of fracture resistance.
The consistency between $A_G$ and $A_\Gamma$ indicates that $2\Gamma$ characterizes the non-equilibrium nature of the fracture.
Specifically, for $\thetaStep<19.11^\circ$, the value $A_G=0.959$ fitted from $\thetaStep-\thetaZZ$ relation matches well with $A_\Gamma$ (Fig.\,\ref{Fig_2}b).
For $\thetaStep>19.11$, the fitting result of $A_G=0.93$ is slightly smaller than $A_\Gamma=0.959$, which is attributed to the nature of asymmetric fracture where cracks advancing along the armchair motif prefer to deflect into the zigzag directions (Fig.\,\ref{Fig_3}b).
The relations between $\thetaStep$ and $\thetaZZ$ summarized in Fig.\,\ref{Fig_5} show that predictions by assuming $A_G=A_{\gamma}$ do not prefer zigzag edges, while the results using $A_G=A_{\Gamma}$ confirm the experimental results \citep{qu_2022}.
We also find that the energy densities of unrelaxed edges are very close to the measured fracture toughness ($\Gc\left(\theta\right) \approx 2\Gamma\left(\theta\right)$) (Figs.\,\ref{Fig_1}b, \ref{Fig_2}b and \ref{Fig_3}), although the strain states at the crack tip are different from that in lattice decohesion \cite{van2004thermodynamics}.

\textit{Conclusion}-
Using a high-fidelity neural network-based force field developed in this work, we find that the kinetics of fracture is much determined by the intermediate, unrelaxed states of the crack tip.
The energy density of relaxed edges widely used in the literature fails to offer a reasonable measure of fracture toughness.
Instead, the $\thetaStep-\thetaZZ$ relation, $\left(\Ktwo/\Kone\right)_{\rm c}$, and $\Gamma\left(\theta\right)$ offer reasonable measures of the fracture anisotropy, that are, $A_G=0.96$, $A_G = 0.935-0.966$, and $A_\Gamma = 0.959$ (for graphene), respectively.
The first two measures can be obtained from experiments or simulations with the atomic-level resolution, while the third one can be considered as material parameters and determined by first-principles calculations.
This work highlights the multiscale and non-equilibrium nature of the fracture and the theory and methodology developed for graphene are extended to other 2D crystals such as h-BN and MoS$_{2}$ (Figs.\,\ref{Fig_4}b-\ref{Fig_4}e)~\cite{supp_info}.

This study was supported by the National Natural Science Foundation of China
through grants 11825203, 11832010, 11921002, and 52090032.
The
computation was performed on the Explorer 100 cluster system of the Tsinghua National Laboratory for Information Science and Technology.

\bibliography{apssamp}

\begin{thebibliography}{60}%
\makeatletter
\providecommand \@ifxundefined [1]{%
 \@ifx{#1\undefined}
}%
\providecommand \@ifnum [1]{%
 \ifnum #1\expandafter \@firstoftwo
 \else \expandafter \@secondoftwo
 \fi
}%
\providecommand \@ifx [1]{%
 \ifx #1\expandafter \@firstoftwo
 \else \expandafter \@secondoftwo
 \fi
}%
\providecommand \natexlab [1]{#1}%
\providecommand \enquote  [1]{``#1''}%
\providecommand \bibnamefont  [1]{#1}%
\providecommand \bibfnamefont [1]{#1}%
\providecommand \citenamefont [1]{#1}%
\providecommand \href@noop [0]{\@secondoftwo}%
\providecommand \href [0]{\begingroup \@sanitize@url \@href}%
\providecommand \@href[1]{\@@startlink{#1}\@@href}%
\providecommand \@@href[1]{\endgroup#1\@@endlink}%
\providecommand \@sanitize@url [0]{\catcode `\\12\catcode `\$12\catcode
  `\&12\catcode `\#12\catcode `\^12\catcode `\_12\catcode `\%12\relax}%
\providecommand \@@startlink[1]{}%
\providecommand \@@endlink[0]{}%
\providecommand \url  [0]{\begingroup\@sanitize@url \@url }%
\providecommand \@url [1]{\endgroup\@href {#1}{\urlprefix }}%
\providecommand \urlprefix  [0]{URL }%
\providecommand \Eprint [0]{\href }%
\providecommand \doibase [0]{https://doi.org/}%
\providecommand \selectlanguage [0]{\@gobble}%
\providecommand \bibinfo  [0]{\@secondoftwo}%
\providecommand \bibfield  [0]{\@secondoftwo}%
\providecommand \translation [1]{[#1]}%
\providecommand \BibitemOpen [0]{}%
\providecommand \bibitemStop [0]{}%
\providecommand \bibitemNoStop [0]{.\EOS\space}%
\providecommand \EOS [0]{\spacefactor3000\relax}%
\providecommand \BibitemShut  [1]{\csname bibitem#1\endcsname}%
\let\auto@bib@innerbib\@empty
\bibitem [{\citenamefont {Feng}\ \emph {et~al.}(2022)\citenamefont {Feng},
  \citenamefont {Cao}, \citenamefont {Gao}, \citenamefont {Han}, \citenamefont
  {Liu}, \citenamefont {Lu},\ and\ \citenamefont {Xu}}]{feng_2022}%
  \BibitemOpen
  \bibfield  {author} {\bibinfo {author} {\bibfnamefont {S.}~\bibnamefont
  {Feng}}, \bibinfo {author} {\bibfnamefont {K.}~\bibnamefont {Cao}}, \bibinfo
  {author} {\bibfnamefont {Y.}~\bibnamefont {Gao}}, \bibinfo {author}
  {\bibfnamefont {Y.}~\bibnamefont {Han}}, \bibinfo {author} {\bibfnamefont
  {Z.}~\bibnamefont {Liu}}, \bibinfo {author} {\bibfnamefont {Y.}~\bibnamefont
  {Lu}},\ and\ \bibinfo {author} {\bibfnamefont {Z.}~\bibnamefont {Xu}},\
  }\bibfield  {title} {\bibinfo {title} {Experimentally measuring weak fracture
  toughness anisotropy in graphene},\ }\href
  {https://doi.org/10.1038/s43246-022-00252-4} {\bibfield  {journal} {\bibinfo
  {journal} {Commun. Mater.}\ }\textbf {\bibinfo {volume} {3}},\ \bibinfo
  {pages} {28} (\bibinfo {year} {2022})}\BibitemShut {NoStop}%
\bibitem [{\citenamefont {Qu}\ \emph {et~al.}(2022)\citenamefont {Qu},
  \citenamefont {Shi}, \citenamefont {Chen}, \citenamefont {Wu}, \citenamefont
  {Wang}, \citenamefont {Shi}, \citenamefont {Gao}, \citenamefont {Xu},\ and\
  \citenamefont {Zheng}}]{qu_2022}%
  \BibitemOpen
  \bibfield  {author} {\bibinfo {author} {\bibfnamefont {C.}~\bibnamefont
  {Qu}}, \bibinfo {author} {\bibfnamefont {D.}~\bibnamefont {Shi}}, \bibinfo
  {author} {\bibfnamefont {L.}~\bibnamefont {Chen}}, \bibinfo {author}
  {\bibfnamefont {Z.}~\bibnamefont {Wu}}, \bibinfo {author} {\bibfnamefont
  {J.}~\bibnamefont {Wang}}, \bibinfo {author} {\bibfnamefont {S.}~\bibnamefont
  {Shi}}, \bibinfo {author} {\bibfnamefont {E.}~\bibnamefont {Gao}}, \bibinfo
  {author} {\bibfnamefont {Z.}~\bibnamefont {Xu}},\ and\ \bibinfo {author}
  {\bibfnamefont {Q.}~\bibnamefont {Zheng}},\ }\bibfield  {title} {\bibinfo
  {title} {Anisotropic fracture of graphene revealed by surface steps on
  graphite},\ }\href {https://doi.org/10.1103/physrevlett.129.026101}
  {\bibfield  {journal} {\bibinfo  {journal} {Phys. Rev. Lett.}\ }\textbf
  {\bibinfo {volume} {129}},\ \bibinfo {pages} {026101} (\bibinfo {year}
  {2022})}\BibitemShut {NoStop}%
\bibitem [{\citenamefont {Fujihara}\ \emph {et~al.}(2015)\citenamefont
  {Fujihara}, \citenamefont {Inoue}, \citenamefont {Kurita}, \citenamefont
  {Taniuchi}, \citenamefont {Motoyui}, \citenamefont {Shin}, \citenamefont
  {Komori}, \citenamefont {Maniwa}, \citenamefont {Shinohara}, \citenamefont
  {Miyata},\ and\ \citenamefont {et~al.}}]{fujihara_2015}%
  \BibitemOpen
  \bibfield  {author} {\bibinfo {author} {\bibfnamefont {M.}~\bibnamefont
  {Fujihara}}, \bibinfo {author} {\bibfnamefont {R.}~\bibnamefont {Inoue}},
  \bibinfo {author} {\bibfnamefont {R.}~\bibnamefont {Kurita}}, \bibinfo
  {author} {\bibfnamefont {T.}~\bibnamefont {Taniuchi}}, \bibinfo {author}
  {\bibfnamefont {Y.}~\bibnamefont {Motoyui}}, \bibinfo {author} {\bibfnamefont
  {S.}~\bibnamefont {Shin}}, \bibinfo {author} {\bibfnamefont {F.}~\bibnamefont
  {Komori}}, \bibinfo {author} {\bibfnamefont {Y.}~\bibnamefont {Maniwa}},
  \bibinfo {author} {\bibfnamefont {H.}~\bibnamefont {Shinohara}}, \bibinfo
  {author} {\bibfnamefont {Y.}~\bibnamefont {Miyata}},\ and\ \bibinfo {author}
  {\bibnamefont {et~al.}},\ }\bibfield  {title} {\bibinfo {title} {Selective
  formation of zigzag edges in graphene cracks},\ }\href
  {https://doi.org/10.1021/acsnano.5b03079} {\bibfield  {journal} {\bibinfo
  {journal} {ACS Nano}\ }\textbf {\bibinfo {volume} {9}},\ \bibinfo {pages}
  {9027} (\bibinfo {year} {2015})}\BibitemShut {NoStop}%
\bibitem [{\citenamefont {Kim}\ \emph {et~al.}(2013)\citenamefont {Kim},
  \citenamefont {Coh}, \citenamefont {Kisielowski}, \citenamefont {Crommie},
  \citenamefont {Louie}, \citenamefont {Cohen},\ and\ \citenamefont
  {Zettl}}]{kim_2013}%
  \BibitemOpen
  \bibfield  {author} {\bibinfo {author} {\bibfnamefont {K.}~\bibnamefont
  {Kim}}, \bibinfo {author} {\bibfnamefont {S.}~\bibnamefont {Coh}}, \bibinfo
  {author} {\bibfnamefont {C.}~\bibnamefont {Kisielowski}}, \bibinfo {author}
  {\bibfnamefont {M.~F.}\ \bibnamefont {Crommie}}, \bibinfo {author}
  {\bibfnamefont {S.~G.}\ \bibnamefont {Louie}}, \bibinfo {author}
  {\bibfnamefont {M.~L.}\ \bibnamefont {Cohen}},\ and\ \bibinfo {author}
  {\bibfnamefont {A.}~\bibnamefont {Zettl}},\ }\bibfield  {title} {\bibinfo
  {title} {Atomically perfect torn graphene edges and their reversible
  reconstruction},\ }\href {https://doi.org/10.1038/ncomms3723} {\bibfield
  {journal} {\bibinfo  {journal} {Nat. Commun.}\ }\textbf {\bibinfo {volume}
  {4}},\ \bibinfo {pages} {2723} (\bibinfo {year} {2013})}\BibitemShut
  {NoStop}%
\bibitem [{\citenamefont {Neubeck}\ \emph {et~al.}(2010)\citenamefont
  {Neubeck}, \citenamefont {You}, \citenamefont {Ni}, \citenamefont {Blake},
  \citenamefont {Shen}, \citenamefont {Geim},\ and\ \citenamefont
  {Novoselov}}]{neubeck_2010}%
  \BibitemOpen
  \bibfield  {author} {\bibinfo {author} {\bibfnamefont {S.}~\bibnamefont
  {Neubeck}}, \bibinfo {author} {\bibfnamefont {Y.~M.}\ \bibnamefont {You}},
  \bibinfo {author} {\bibfnamefont {Z.~H.}\ \bibnamefont {Ni}}, \bibinfo
  {author} {\bibfnamefont {P.}~\bibnamefont {Blake}}, \bibinfo {author}
  {\bibfnamefont {Z.~X.}\ \bibnamefont {Shen}}, \bibinfo {author}
  {\bibfnamefont {A.~K.}\ \bibnamefont {Geim}},\ and\ \bibinfo {author}
  {\bibfnamefont {K.~S.}\ \bibnamefont {Novoselov}},\ }\bibfield  {title}
  {\bibinfo {title} {Direct determination of the crystallographic orientation
  of graphene edges by atomic resolution imaging},\ }\href
  {https://doi.org/10.1063/1.3467468} {\bibfield  {journal} {\bibinfo
  {journal} {Appl. Phys. Lett.}\ }\textbf {\bibinfo {volume} {97}},\ \bibinfo
  {pages} {053110} (\bibinfo {year} {2010})}\BibitemShut {NoStop}%
\bibitem [{\citenamefont {Jia}\ \emph {et~al.}(2009)\citenamefont {Jia},
  \citenamefont {Hofmann}, \citenamefont {Meunier}, \citenamefont {Sumpter},
  \citenamefont {Campos-Delgado}, \citenamefont {Romo-Herrera}, \citenamefont
  {Son}, \citenamefont {Hsieh}, \citenamefont {Reina}, \citenamefont {Kong},\
  and\ \citenamefont {et~al.}}]{jia_2009}%
  \BibitemOpen
  \bibfield  {author} {\bibinfo {author} {\bibfnamefont {X.}~\bibnamefont
  {Jia}}, \bibinfo {author} {\bibfnamefont {M.}~\bibnamefont {Hofmann}},
  \bibinfo {author} {\bibfnamefont {V.}~\bibnamefont {Meunier}}, \bibinfo
  {author} {\bibfnamefont {B.~G.}\ \bibnamefont {Sumpter}}, \bibinfo {author}
  {\bibfnamefont {J.}~\bibnamefont {Campos-Delgado}}, \bibinfo {author}
  {\bibfnamefont {J.~M.}\ \bibnamefont {Romo-Herrera}}, \bibinfo {author}
  {\bibfnamefont {H.}~\bibnamefont {Son}}, \bibinfo {author} {\bibfnamefont
  {Y.-P.}\ \bibnamefont {Hsieh}}, \bibinfo {author} {\bibfnamefont
  {A.}~\bibnamefont {Reina}}, \bibinfo {author} {\bibfnamefont
  {J.}~\bibnamefont {Kong}},\ and\ \bibinfo {author} {\bibnamefont {et~al.}},\
  }\bibfield  {title} {\bibinfo {title} {Controlled formation of sharp zigzag
  and armchair edges in graphitic nanoribbons},\ }\href
  {https://doi.org/10.1126/science.1166862} {\bibfield  {journal} {\bibinfo
  {journal} {Science}\ }\textbf {\bibinfo {volume} {323}},\ \bibinfo {pages}
  {1701} (\bibinfo {year} {2009})}\BibitemShut {NoStop}%
\bibitem [{\citenamefont {Shi}\ \emph {et~al.}(2020)\citenamefont {Shi},
  \citenamefont {Yang}, \citenamefont {Deng}, \citenamefont {Tong},
  \citenamefont {Wu}, \citenamefont {Zhang}, \citenamefont {Zhang},
  \citenamefont {Yin},\ and\ \citenamefont {Qin}}]{shi_2020}%
  \BibitemOpen
  \bibfield  {author} {\bibinfo {author} {\bibfnamefont {L.-J.}\ \bibnamefont
  {Shi}}, \bibinfo {author} {\bibfnamefont {L.-Z.}\ \bibnamefont {Yang}},
  \bibinfo {author} {\bibfnamefont {J.-Q.}\ \bibnamefont {Deng}}, \bibinfo
  {author} {\bibfnamefont {L.-H.}\ \bibnamefont {Tong}}, \bibinfo {author}
  {\bibfnamefont {Q.}~\bibnamefont {Wu}}, \bibinfo {author} {\bibfnamefont
  {L.}~\bibnamefont {Zhang}}, \bibinfo {author} {\bibfnamefont
  {L.}~\bibnamefont {Zhang}}, \bibinfo {author} {\bibfnamefont {L.-J.}\
  \bibnamefont {Yin}},\ and\ \bibinfo {author} {\bibfnamefont {Z.}~\bibnamefont
  {Qin}},\ }\bibfield  {title} {\bibinfo {title} {Constructing graphene
  nanostructures with zigzag edge terminations by controllable {STM} tearing
  and folding},\ }\href {https://doi.org/10.1016/j.carbon.2020.04.070}
  {\bibfield  {journal} {\bibinfo  {journal} {Carbon}\ }\textbf {\bibinfo
  {volume} {165}},\ \bibinfo {pages} {169} (\bibinfo {year}
  {2020})}\BibitemShut {NoStop}%
\bibitem [{\citenamefont {Girit}\ \emph {et~al.}(2009)\citenamefont {Girit},
  \citenamefont {Meyer}, \citenamefont {Erni}, \citenamefont {Rossell},
  \citenamefont {Kisielowski}, \citenamefont {Yang}, \citenamefont {Park},
  \citenamefont {Crommie}, \citenamefont {Cohen}, \citenamefont {Louie} \emph
  {et~al.}}]{girit_2009}%
  \BibitemOpen
  \bibfield  {author} {\bibinfo {author} {\bibfnamefont {{\c{C}}.~O.}\
  \bibnamefont {Girit}}, \bibinfo {author} {\bibfnamefont {J.~C.}\ \bibnamefont
  {Meyer}}, \bibinfo {author} {\bibfnamefont {R.}~\bibnamefont {Erni}},
  \bibinfo {author} {\bibfnamefont {M.~D.}\ \bibnamefont {Rossell}}, \bibinfo
  {author} {\bibfnamefont {C.}~\bibnamefont {Kisielowski}}, \bibinfo {author}
  {\bibfnamefont {L.}~\bibnamefont {Yang}}, \bibinfo {author} {\bibfnamefont
  {C.-H.}\ \bibnamefont {Park}}, \bibinfo {author} {\bibfnamefont
  {M.}~\bibnamefont {Crommie}}, \bibinfo {author} {\bibfnamefont {M.~L.}\
  \bibnamefont {Cohen}}, \bibinfo {author} {\bibfnamefont {S.~G.}\ \bibnamefont
  {Louie}}, \emph {et~al.},\ }\bibfield  {title} {\bibinfo {title} {Graphene at
  the edge: {S}tability and dynamics},\ }\href
  {https://doi.org/10.1126/science.1166999} {\bibfield  {journal} {\bibinfo
  {journal} {Science}\ }\textbf {\bibinfo {volume} {323}},\ \bibinfo {pages}
  {1705} (\bibinfo {year} {2009})}\BibitemShut {NoStop}%
\bibitem [{\citenamefont {Kim}\ \emph {et~al.}(2012)\citenamefont {Kim},
  \citenamefont {Artyukhov}, \citenamefont {Regan}, \citenamefont {Liu},
  \citenamefont {Crommie}, \citenamefont {Yakobson},\ and\ \citenamefont
  {Zettl}}]{kim_2012}%
  \BibitemOpen
  \bibfield  {author} {\bibinfo {author} {\bibfnamefont {K.}~\bibnamefont
  {Kim}}, \bibinfo {author} {\bibfnamefont {V.~I.}\ \bibnamefont {Artyukhov}},
  \bibinfo {author} {\bibfnamefont {W.}~\bibnamefont {Regan}}, \bibinfo
  {author} {\bibfnamefont {Y.}~\bibnamefont {Liu}}, \bibinfo {author}
  {\bibfnamefont {M.~F.}\ \bibnamefont {Crommie}}, \bibinfo {author}
  {\bibfnamefont {B.~I.}\ \bibnamefont {Yakobson}},\ and\ \bibinfo {author}
  {\bibfnamefont {A.}~\bibnamefont {Zettl}},\ }\bibfield  {title} {\bibinfo
  {title} {Ripping graphene: {P}referred directions},\ }\href
  {https://doi.org/10.1021/nl203547z} {\bibfield  {journal} {\bibinfo
  {journal} {Nano Lett.}\ }\textbf {\bibinfo {volume} {12}},\ \bibinfo {pages}
  {293} (\bibinfo {year} {2012})}\BibitemShut {NoStop}%
\bibitem [{\citenamefont {Jun}(2008)}]{jun_2008}%
  \BibitemOpen
  \bibfield  {author} {\bibinfo {author} {\bibfnamefont {S.}~\bibnamefont
  {Jun}},\ }\bibfield  {title} {\bibinfo {title} {Density-functional study of
  edge stress in graphene},\ }\href
  {https://doi.org/10.1103/physrevb.78.073405} {\bibfield  {journal} {\bibinfo
  {journal} {Phys. Rev. B}\ }\textbf {\bibinfo {volume} {78}},\ \bibinfo
  {pages} {073405} (\bibinfo {year} {2008})}\BibitemShut {NoStop}%
\bibitem [{\citenamefont {Koskinen}\ \emph {et~al.}(2008)\citenamefont
  {Koskinen}, \citenamefont {Malola},\ and\ \citenamefont
  {Häkkinen}}]{koskinen_2008}%
  \BibitemOpen
  \bibfield  {author} {\bibinfo {author} {\bibfnamefont {P.}~\bibnamefont
  {Koskinen}}, \bibinfo {author} {\bibfnamefont {S.}~\bibnamefont {Malola}},\
  and\ \bibinfo {author} {\bibfnamefont {H.}~\bibnamefont {Häkkinen}},\
  }\bibfield  {title} {\bibinfo {title} {Self-passivating edge reconstructions
  of graphene},\ }\href {https://doi.org/10.1103/physrevlett.101.115502}
  {\bibfield  {journal} {\bibinfo  {journal} {Phys. Rev. Lett.}\ }\textbf
  {\bibinfo {volume} {101}},\ \bibinfo {pages} {115502} (\bibinfo {year}
  {2008})}\BibitemShut {NoStop}%
\bibitem [{\citenamefont {Huang}\ \emph {et~al.}(2009)\citenamefont {Huang},
  \citenamefont {Liu}, \citenamefont {Su}, \citenamefont {Wu}, \citenamefont
  {Duan}, \citenamefont {Gu},\ and\ \citenamefont {Liu}}]{huang_2009}%
  \BibitemOpen
  \bibfield  {author} {\bibinfo {author} {\bibfnamefont {B.}~\bibnamefont
  {Huang}}, \bibinfo {author} {\bibfnamefont {M.}~\bibnamefont {Liu}}, \bibinfo
  {author} {\bibfnamefont {N.}~\bibnamefont {Su}}, \bibinfo {author}
  {\bibfnamefont {J.}~\bibnamefont {Wu}}, \bibinfo {author} {\bibfnamefont
  {W.}~\bibnamefont {Duan}}, \bibinfo {author} {\bibfnamefont {B.-l.}\
  \bibnamefont {Gu}},\ and\ \bibinfo {author} {\bibfnamefont {F.}~\bibnamefont
  {Liu}},\ }\bibfield  {title} {\bibinfo {title} {Quantum manifestations of
  graphene edge stress and edge instability: {A} first-principles study},\
  }\href {https://doi.org/10.1103/physrevlett.102.166404} {\bibfield  {journal}
  {\bibinfo  {journal} {Phys. Rev. Lett.}\ }\textbf {\bibinfo {volume} {102}},\
  \bibinfo {pages} {166404} (\bibinfo {year} {2009})}\BibitemShut {NoStop}%
\bibitem [{\citenamefont {Gan}\ and\ \citenamefont
  {Srolovitz}(2010)}]{gan_2010}%
  \BibitemOpen
  \bibfield  {author} {\bibinfo {author} {\bibfnamefont {C.~K.}\ \bibnamefont
  {Gan}}\ and\ \bibinfo {author} {\bibfnamefont {D.~J.}\ \bibnamefont
  {Srolovitz}},\ }\bibfield  {title} {\bibinfo {title} {First-principles study
  of graphene edge properties and flake shapes},\ }\href
  {https://doi.org/10.1103/physrevb.81.125445} {\bibfield  {journal} {\bibinfo
  {journal} {Phys. Rev. B}\ }\textbf {\bibinfo {volume} {81}},\ \bibinfo
  {pages} {125445} (\bibinfo {year} {2010})}\BibitemShut {NoStop}%
\bibitem [{\citenamefont {Liu}\ \emph {et~al.}(2010)\citenamefont {Liu},
  \citenamefont {Dobrinsky},\ and\ \citenamefont {Yakobson}}]{Boris_2010}%
  \BibitemOpen
  \bibfield  {author} {\bibinfo {author} {\bibfnamefont {Y.}~\bibnamefont
  {Liu}}, \bibinfo {author} {\bibfnamefont {A.}~\bibnamefont {Dobrinsky}},\
  and\ \bibinfo {author} {\bibfnamefont {B.~I.}\ \bibnamefont {Yakobson}},\
  }\bibfield  {title} {\bibinfo {title} {Graphene edge from armchair to zigzag:
  {T}he origins of nanotube chirality?},\ }\href
  {https://doi.org/10.1103/physrevlett.105.235502} {\bibfield  {journal}
  {\bibinfo  {journal} {Phys. Rev. Lett.}\ }\textbf {\bibinfo {volume} {105}},\
  \bibinfo {pages} {235502} (\bibinfo {year} {2010})}\BibitemShut {NoStop}%
\bibitem [{\citenamefont {Gao}\ \emph {et~al.}(2011)\citenamefont {Gao},
  \citenamefont {Yip}, \citenamefont {Zhao}, \citenamefont {Yakobson},\ and\
  \citenamefont {Ding}}]{gao_2011}%
  \BibitemOpen
  \bibfield  {author} {\bibinfo {author} {\bibfnamefont {J.}~\bibnamefont
  {Gao}}, \bibinfo {author} {\bibfnamefont {J.}~\bibnamefont {Yip}}, \bibinfo
  {author} {\bibfnamefont {J.}~\bibnamefont {Zhao}}, \bibinfo {author}
  {\bibfnamefont {B.~I.}\ \bibnamefont {Yakobson}},\ and\ \bibinfo {author}
  {\bibfnamefont {F.}~\bibnamefont {Ding}},\ }\bibfield  {title} {\bibinfo
  {title} {Graphene nucleation on transition metal surface: {S}tructure
  transformation and role of the metal step edge},\ }\href
  {https://doi.org/10.1021/ja110927p} {\bibfield  {journal} {\bibinfo
  {journal} {J. Am. Chem. Soc.}\ }\textbf {\bibinfo {volume} {133}},\ \bibinfo
  {pages} {5009} (\bibinfo {year} {2011})}\BibitemShut {NoStop}%
\bibitem [{\citenamefont {Yin}\ \emph {et~al.}(2015)\citenamefont {Yin},
  \citenamefont {Qi}, \citenamefont {Fan}, \citenamefont {Zhu}, \citenamefont
  {Wang},\ and\ \citenamefont {Wei}}]{yin_2015}%
  \BibitemOpen
  \bibfield  {author} {\bibinfo {author} {\bibfnamefont {H.}~\bibnamefont
  {Yin}}, \bibinfo {author} {\bibfnamefont {H.~J.}\ \bibnamefont {Qi}},
  \bibinfo {author} {\bibfnamefont {F.}~\bibnamefont {Fan}}, \bibinfo {author}
  {\bibfnamefont {T.}~\bibnamefont {Zhu}}, \bibinfo {author} {\bibfnamefont
  {B.}~\bibnamefont {Wang}},\ and\ \bibinfo {author} {\bibfnamefont
  {Y.}~\bibnamefont {Wei}},\ }\bibfield  {title} {\bibinfo {title} {Griffith
  criterion for brittle fracture in graphene},\ }\href
  {https://doi.org/10.1021/nl5047686} {\bibfield  {journal} {\bibinfo
  {journal} {Nano Lett.}\ }\textbf {\bibinfo {volume} {15}},\ \bibinfo {pages}
  {1918} (\bibinfo {year} {2015})}\BibitemShut {NoStop}%
\bibitem [{sup()}]{supp_info}%
  \BibitemOpen
  \href@noop {} {\bibinfo {title} {{See Supplemental Material at
  http://link.aps.org/ supplemental/10.1103/PhysRevLett.xxx.xxxxxx, which
  includes detailed methods, derivations of the theoretical model, supplemental
  figures and tables, datasets and
  Refs.\cite{lammps,dpgen,deepmd,handbook,over-determined,
  wilson_2019,williams1957,feng_2022,nuismer1975,siesta,PBE,troullier1991,VASP,PAW,kim_2012,huang_2009,zhang_2014,wang_2016,Boris_2010,koskinen_2008,gan_2010,addou_2018,xiao_2016,lee_2017,aghaei_2016,zhou_2013,yang_2017,AFM_2021,TEM_1993,dataset}.}}}\BibitemShut
  {Stop}%
\bibitem [{\citenamefont {Zhang}\ \emph {et~al.}(2014)\citenamefont {Zhang},
  \citenamefont {Ma}, \citenamefont {Fan}, \citenamefont {Zeng}, \citenamefont
  {Peng}, \citenamefont {Loya}, \citenamefont {Liu}, \citenamefont {Gong},
  \citenamefont {Zhang}, \citenamefont {Zhang},\ and\ \citenamefont
  {et~al.}}]{zhang_2014}%
  \BibitemOpen
  \bibfield  {author} {\bibinfo {author} {\bibfnamefont {P.}~\bibnamefont
  {Zhang}}, \bibinfo {author} {\bibfnamefont {L.}~\bibnamefont {Ma}}, \bibinfo
  {author} {\bibfnamefont {F.}~\bibnamefont {Fan}}, \bibinfo {author}
  {\bibfnamefont {Z.}~\bibnamefont {Zeng}}, \bibinfo {author} {\bibfnamefont
  {C.}~\bibnamefont {Peng}}, \bibinfo {author} {\bibfnamefont {P.~E.}\
  \bibnamefont {Loya}}, \bibinfo {author} {\bibfnamefont {Z.}~\bibnamefont
  {Liu}}, \bibinfo {author} {\bibfnamefont {Y.}~\bibnamefont {Gong}}, \bibinfo
  {author} {\bibfnamefont {J.}~\bibnamefont {Zhang}}, \bibinfo {author}
  {\bibfnamefont {X.}~\bibnamefont {Zhang}},\ and\ \bibinfo {author}
  {\bibnamefont {et~al.}},\ }\bibfield  {title} {\bibinfo {title} {Fracture
  toughness of graphene},\ }\href {https://doi.org/10.1038/ncomms4782}
  {\bibfield  {journal} {\bibinfo  {journal} {Nat. Commun.}\ }\textbf {\bibinfo
  {volume} {5}},\ \bibinfo {pages} {3782} (\bibinfo {year} {2014})}\BibitemShut
  {NoStop}%
\bibitem [{\citenamefont {Sen}\ \emph {et~al.}(2010)\citenamefont {Sen},
  \citenamefont {Novoselov}, \citenamefont {Reis},\ and\ \citenamefont
  {Buehler}}]{sen_2010}%
  \BibitemOpen
  \bibfield  {author} {\bibinfo {author} {\bibfnamefont {D.}~\bibnamefont
  {Sen}}, \bibinfo {author} {\bibfnamefont {K.~S.}\ \bibnamefont {Novoselov}},
  \bibinfo {author} {\bibfnamefont {P.~M.}\ \bibnamefont {Reis}},\ and\
  \bibinfo {author} {\bibfnamefont {M.~J.}\ \bibnamefont {Buehler}},\
  }\bibfield  {title} {\bibinfo {title} {Tearing graphene sheets from adhesive
  substrates produces tapered nanoribbons},\ }\href
  {https://doi.org/10.1002/smll.201000097} {\bibfield  {journal} {\bibinfo
  {journal} {Small}\ }\textbf {\bibinfo {volume} {6}},\ \bibinfo {pages} {1108}
  (\bibinfo {year} {2010})}\BibitemShut {NoStop}%
\bibitem [{\citenamefont {Song}\ \emph {et~al.}(2017)\citenamefont {Song},
  \citenamefont {Ni},\ and\ \citenamefont {Xu}}]{song_2017}%
  \BibitemOpen
  \bibfield  {author} {\bibinfo {author} {\bibfnamefont {Z.}~\bibnamefont
  {Song}}, \bibinfo {author} {\bibfnamefont {Y.}~\bibnamefont {Ni}},\ and\
  \bibinfo {author} {\bibfnamefont {Z.}~\bibnamefont {Xu}},\ }\bibfield
  {title} {\bibinfo {title} {Geometrical distortion leads to {G}riffith
  strength reduction in graphene membranes},\ }\href
  {https://doi.org/10.1016/j.eml.2017.01.005} {\bibfield  {journal} {\bibinfo
  {journal} {Extreme Mech. Lett.}\ }\textbf {\bibinfo {volume} {14}},\ \bibinfo
  {pages} {31} (\bibinfo {year} {2017})}\BibitemShut {NoStop}%
\bibitem [{\citenamefont {Lawn}(1993)}]{lawn_2004}%
  \BibitemOpen
  \bibfield  {author} {\bibinfo {author} {\bibfnamefont {B.~R.}\ \bibnamefont
  {Lawn}},\ }\href@noop {} {\emph {\bibinfo {title} {Fracture of {Brittle}
  {Solids}}}}\ (\bibinfo  {publisher} {Cambridge University Press},\ \bibinfo
  {year} {1993})\BibitemShut {NoStop}%
\bibitem [{\citenamefont {Griffith}(1921)}]{griffith_1921}%
  \BibitemOpen
  \bibfield  {author} {\bibinfo {author} {\bibfnamefont {A.~A.}\ \bibnamefont
  {Griffith}},\ }\bibfield  {title} {\bibinfo {title} {{VI}. {The} phenomena of
  rupture and flow in solids},\ }\href {https://doi.org/10.1098/rsta.1921.0006}
  {\bibfield  {journal} {\bibinfo  {journal} {Philos. Trans. R. Soc. A}\
  }\textbf {\bibinfo {volume} {221}},\ \bibinfo {pages} {163} (\bibinfo {year}
  {1921})}\BibitemShut {NoStop}%
\bibitem [{\citenamefont {Yang}\ \emph {et~al.}(2021)\citenamefont {Yang},
  \citenamefont {Song}, \citenamefont {Lu}, \citenamefont {Zhang},
  \citenamefont {Zhang}, \citenamefont {Ni}, \citenamefont {Wang},
  \citenamefont {Li}, \citenamefont {Gu}, \citenamefont {Xie},\ and\
  \citenamefont {et~al.}}]{zhigong_2021}%
  \BibitemOpen
  \bibfield  {author} {\bibinfo {author} {\bibfnamefont {Y.}~\bibnamefont
  {Yang}}, \bibinfo {author} {\bibfnamefont {Z.}~\bibnamefont {Song}}, \bibinfo
  {author} {\bibfnamefont {G.}~\bibnamefont {Lu}}, \bibinfo {author}
  {\bibfnamefont {Q.}~\bibnamefont {Zhang}}, \bibinfo {author} {\bibfnamefont
  {B.}~\bibnamefont {Zhang}}, \bibinfo {author} {\bibfnamefont
  {B.}~\bibnamefont {Ni}}, \bibinfo {author} {\bibfnamefont {C.}~\bibnamefont
  {Wang}}, \bibinfo {author} {\bibfnamefont {X.}~\bibnamefont {Li}}, \bibinfo
  {author} {\bibfnamefont {L.}~\bibnamefont {Gu}}, \bibinfo {author}
  {\bibfnamefont {X.}~\bibnamefont {Xie}},\ and\ \bibinfo {author}
  {\bibnamefont {et~al.}},\ }\bibfield  {title} {\bibinfo {title} {Intrinsic
  toughening and stable crack propagation in hexagonal boron nitride},\ }\href
  {https://doi.org/10.1038/s41586-021-03488-1} {\bibfield  {journal} {\bibinfo
  {journal} {Nature}\ }\textbf {\bibinfo {volume} {594}},\ \bibinfo {pages}
  {57} (\bibinfo {year} {2021})}\BibitemShut {NoStop}%
\bibitem [{\citenamefont {Takei}\ \emph {et~al.}(2013)\citenamefont {Takei},
  \citenamefont {Roman}, \citenamefont {Bico}, \citenamefont {Hamm},\ and\
  \citenamefont {Melo}}]{takei_2013}%
  \BibitemOpen
  \bibfield  {author} {\bibinfo {author} {\bibfnamefont {A.}~\bibnamefont
  {Takei}}, \bibinfo {author} {\bibfnamefont {B.}~\bibnamefont {Roman}},
  \bibinfo {author} {\bibfnamefont {J.}~\bibnamefont {Bico}}, \bibinfo {author}
  {\bibfnamefont {E.}~\bibnamefont {Hamm}},\ and\ \bibinfo {author}
  {\bibfnamefont {F.}~\bibnamefont {Melo}},\ }\bibfield  {title} {\bibinfo
  {title} {Forbidden directions for the fracture of thin anisotropic sheets:
  {A}n analogy with the {Wulff} plot},\ }\href
  {https://doi.org/10.1103/physrevlett.110.144301} {\bibfield  {journal}
  {\bibinfo  {journal} {Phys. Rev. Lett.}\ }\textbf {\bibinfo {volume} {110}},\
  \bibinfo {pages} {144301} (\bibinfo {year} {2013})}\BibitemShut {NoStop}%
\bibitem [{\citenamefont {DelRio}\ \emph {et~al.}(2015)\citenamefont {DelRio},
  \citenamefont {Cook},\ and\ \citenamefont {Boyce}}]{delrio_2015}%
  \BibitemOpen
  \bibfield  {author} {\bibinfo {author} {\bibfnamefont {F.~W.}\ \bibnamefont
  {DelRio}}, \bibinfo {author} {\bibfnamefont {R.~F.}\ \bibnamefont {Cook}},\
  and\ \bibinfo {author} {\bibfnamefont {B.~L.}\ \bibnamefont {Boyce}},\
  }\bibfield  {title} {\bibinfo {title} {Fracture strength of micro-and
  nano-scale silicon components},\ }\href {https://doi.org/10.1063/1.4919540}
  {\bibfield  {journal} {\bibinfo  {journal} {Appl. Phys. Rev.}\ }\textbf
  {\bibinfo {volume} {2}},\ \bibinfo {pages} {021303} (\bibinfo {year}
  {2015})}\BibitemShut {NoStop}%
\bibitem [{\citenamefont {DelRio}\ \emph {et~al.}(2022)\citenamefont {DelRio},
  \citenamefont {Grutzik}, \citenamefont {Mook}, \citenamefont {Dickens},
  \citenamefont {Kotula}, \citenamefont {Hintsala}, \citenamefont {Stauffer},\
  and\ \citenamefont {Boyce}}]{delrio_2022}%
  \BibitemOpen
  \bibfield  {author} {\bibinfo {author} {\bibfnamefont {F.~W.}\ \bibnamefont
  {DelRio}}, \bibinfo {author} {\bibfnamefont {S.~J.}\ \bibnamefont {Grutzik}},
  \bibinfo {author} {\bibfnamefont {W.~M.}\ \bibnamefont {Mook}}, \bibinfo
  {author} {\bibfnamefont {S.~M.}\ \bibnamefont {Dickens}}, \bibinfo {author}
  {\bibfnamefont {P.~G.}\ \bibnamefont {Kotula}}, \bibinfo {author}
  {\bibfnamefont {E.~D.}\ \bibnamefont {Hintsala}}, \bibinfo {author}
  {\bibfnamefont {D.~D.}\ \bibnamefont {Stauffer}},\ and\ \bibinfo {author}
  {\bibfnamefont {B.~L.}\ \bibnamefont {Boyce}},\ }\bibfield  {title} {\bibinfo
  {title} {Eliciting stable nanoscale fracture in single-crystal silicon},\
  }\href {https://doi.org/10.1080/21663831.2022.2088251} {\bibfield  {journal}
  {\bibinfo  {journal} {Mater. Res. Lett.}\ }\textbf {\bibinfo {volume} {10}},\
  \bibinfo {pages} {728} (\bibinfo {year} {2022})}\BibitemShut {NoStop}%
\bibitem [{\citenamefont {Jung}\ \emph {et~al.}(2019)\citenamefont {Jung},
  \citenamefont {Wang}, \citenamefont {Qin}, \citenamefont {Zhou},
  \citenamefont {Danaie}, \citenamefont {Kirkland}, \citenamefont {Buehler},\
  and\ \citenamefont {Warner}}]{jung_2019}%
  \BibitemOpen
  \bibfield  {author} {\bibinfo {author} {\bibfnamefont {G.~S.}\ \bibnamefont
  {Jung}}, \bibinfo {author} {\bibfnamefont {S.}~\bibnamefont {Wang}}, \bibinfo
  {author} {\bibfnamefont {Z.}~\bibnamefont {Qin}}, \bibinfo {author}
  {\bibfnamefont {S.}~\bibnamefont {Zhou}}, \bibinfo {author} {\bibfnamefont
  {M.}~\bibnamefont {Danaie}}, \bibinfo {author} {\bibfnamefont {A.~I.}\
  \bibnamefont {Kirkland}}, \bibinfo {author} {\bibfnamefont {M.~J.}\
  \bibnamefont {Buehler}},\ and\ \bibinfo {author} {\bibfnamefont {J.~H.}\
  \bibnamefont {Warner}},\ }\bibfield  {title} {\bibinfo {title} {Anisotropic
  fracture dynamics due to local lattice distortions},\ }\href
  {https://doi.org/10.1021/acsnano.9b01071} {\bibfield  {journal} {\bibinfo
  {journal} {ACS Nano}\ }\textbf {\bibinfo {volume} {13}},\ \bibinfo {pages}
  {5693} (\bibinfo {year} {2019})}\BibitemShut {NoStop}%
\bibitem [{\citenamefont {Zhang}\ \emph {et~al.}(2022)\citenamefont {Zhang},
  \citenamefont {Nguyen}, \citenamefont {Zhang}, \citenamefont {Ajayan},
  \citenamefont {Wen},\ and\ \citenamefont {Espinosa}}]{zhang2022atomistic}%
  \BibitemOpen
  \bibfield  {author} {\bibinfo {author} {\bibfnamefont {X.}~\bibnamefont
  {Zhang}}, \bibinfo {author} {\bibfnamefont {H.}~\bibnamefont {Nguyen}},
  \bibinfo {author} {\bibfnamefont {X.}~\bibnamefont {Zhang}}, \bibinfo
  {author} {\bibfnamefont {P.~M.}\ \bibnamefont {Ajayan}}, \bibinfo {author}
  {\bibfnamefont {J.}~\bibnamefont {Wen}},\ and\ \bibinfo {author}
  {\bibfnamefont {H.~D.}\ \bibnamefont {Espinosa}},\ }\bibfield  {title}
  {\bibinfo {title} {Atomistic measurement and modeling of intrinsic fracture
  toughness of two-dimensional materials},\ }\href
  {https://doi.org/10.1073/pnas.2206756119} {\bibfield  {journal} {\bibinfo
  {journal} {Proc. Natl. Acad. Sci.}\ }\textbf {\bibinfo {volume} {119}},\
  \bibinfo {pages} {e2206756119} (\bibinfo {year} {2022})}\BibitemShut
  {NoStop}%
\bibitem [{\citenamefont {Hossain}\ \emph {et~al.}(2018)\citenamefont
  {Hossain}, \citenamefont {Ahmed}, \citenamefont {Silverman}, \citenamefont
  {Khawaja}, \citenamefont {Calderon}, \citenamefont {Rutten},\ and\
  \citenamefont {Tse}}]{hossain_2018}%
  \BibitemOpen
  \bibfield  {author} {\bibinfo {author} {\bibfnamefont {M.~Z.}\ \bibnamefont
  {Hossain}}, \bibinfo {author} {\bibfnamefont {T.}~\bibnamefont {Ahmed}},
  \bibinfo {author} {\bibfnamefont {B.}~\bibnamefont {Silverman}}, \bibinfo
  {author} {\bibfnamefont {M.~S.}\ \bibnamefont {Khawaja}}, \bibinfo {author}
  {\bibfnamefont {J.}~\bibnamefont {Calderon}}, \bibinfo {author}
  {\bibfnamefont {A.}~\bibnamefont {Rutten}},\ and\ \bibinfo {author}
  {\bibfnamefont {S.}~\bibnamefont {Tse}},\ }\bibfield  {title} {\bibinfo
  {title} {Anisotropic toughness and strength in graphene and its atomistic
  origin},\ }\href {https://doi.org/10.1016/j.jmps.2017.09.012} {\bibfield
  {journal} {\bibinfo  {journal} {J. Mech. Phys. Solids}\ }\textbf {\bibinfo
  {volume} {110}},\ \bibinfo {pages} {118} (\bibinfo {year}
  {2018})}\BibitemShut {NoStop}%
\bibitem [{\citenamefont {Friederich}\ \emph {et~al.}(2021)\citenamefont
  {Friederich}, \citenamefont {Häse}, \citenamefont {Proppe},\ and\
  \citenamefont {Aspuru-Guzik}}]{friederich_2021}%
  \BibitemOpen
  \bibfield  {author} {\bibinfo {author} {\bibfnamefont {P.}~\bibnamefont
  {Friederich}}, \bibinfo {author} {\bibfnamefont {F.}~\bibnamefont {Häse}},
  \bibinfo {author} {\bibfnamefont {J.}~\bibnamefont {Proppe}},\ and\ \bibinfo
  {author} {\bibfnamefont {A.}~\bibnamefont {Aspuru-Guzik}},\ }\bibfield
  {title} {\bibinfo {title} {Machine-learned potentials for next-generation
  matter simulations},\ }\href {https://doi.org/10.1038/s41563-020-0777-6}
  {\bibfield  {journal} {\bibinfo  {journal} {Nat. Mater.}\ }\textbf {\bibinfo
  {volume} {20}},\ \bibinfo {pages} {750} (\bibinfo {year} {2021})}\BibitemShut
  {NoStop}%
\bibitem [{\citenamefont {Galib}\ and\ \citenamefont
  {Limmer}(2021)}]{galib2021reactive}%
  \BibitemOpen
  \bibfield  {author} {\bibinfo {author} {\bibfnamefont {M.}~\bibnamefont
  {Galib}}\ and\ \bibinfo {author} {\bibfnamefont {D.~T.}\ \bibnamefont
  {Limmer}},\ }\bibfield  {title} {\bibinfo {title} {Reactive uptake of
  {$N_2O_5$} by atmospheric aerosol is dominated by interfacial processes},\
  }\href {https://doi.org/10.1126/science.abd771} {\bibfield  {journal}
  {\bibinfo  {journal} {Science}\ }\textbf {\bibinfo {volume} {371}},\ \bibinfo
  {pages} {921} (\bibinfo {year} {2021})}\BibitemShut {NoStop}%
\bibitem [{\citenamefont {Font-Clos}\ \emph {et~al.}(2022)\citenamefont
  {Font-Clos}, \citenamefont {Zanchi}, \citenamefont {Hiemer}, \citenamefont
  {Bonfanti}, \citenamefont {Guerra}, \citenamefont {Zaiser},\ and\
  \citenamefont {Zapperi}}]{font2022predicting}%
  \BibitemOpen
  \bibfield  {author} {\bibinfo {author} {\bibfnamefont {F.}~\bibnamefont
  {Font-Clos}}, \bibinfo {author} {\bibfnamefont {M.}~\bibnamefont {Zanchi}},
  \bibinfo {author} {\bibfnamefont {S.}~\bibnamefont {Hiemer}}, \bibinfo
  {author} {\bibfnamefont {S.}~\bibnamefont {Bonfanti}}, \bibinfo {author}
  {\bibfnamefont {R.}~\bibnamefont {Guerra}}, \bibinfo {author} {\bibfnamefont
  {M.}~\bibnamefont {Zaiser}},\ and\ \bibinfo {author} {\bibfnamefont
  {S.}~\bibnamefont {Zapperi}},\ }\bibfield  {title} {\bibinfo {title}
  {Predicting the failure of two-dimensional silica glasses},\ }\href
  {https://doi.org/10.1038/s41467-022-30530-1} {\bibfield  {journal} {\bibinfo
  {journal} {Nat. Commun.}\ }\textbf {\bibinfo {volume} {13}},\ \bibinfo
  {pages} {2820} (\bibinfo {year} {2022})}\BibitemShut {NoStop}%
\bibitem [{\citenamefont {Li}\ and\ \citenamefont {Ding}(2022)}]{li2022origin}%
  \BibitemOpen
  \bibfield  {author} {\bibinfo {author} {\bibfnamefont {P.}~\bibnamefont
  {Li}}\ and\ \bibinfo {author} {\bibfnamefont {F.}~\bibnamefont {Ding}},\
  }\bibfield  {title} {\bibinfo {title} {Origin of the herringbone
  reconstruction of {Au} (111) surface at the atomic scale},\ }\href
  {https://doi.org/10.1126/sciadv.abq2900} {\bibfield  {journal} {\bibinfo
  {journal} {Science Advances}\ }\textbf {\bibinfo {volume} {8}},\ \bibinfo
  {pages} {eabq2900} (\bibinfo {year} {2022})}\BibitemShut {NoStop}%
\bibitem [{\citenamefont {Hedman}\ \emph {et~al.}(2023)\citenamefont {Hedman},
  \citenamefont {McLean}, \citenamefont {Bichara}, \citenamefont {Maruyama},
  \citenamefont {Larsson},\ and\ \citenamefont {Ding}}]{hedman2023dynamics}%
  \BibitemOpen
  \bibfield  {author} {\bibinfo {author} {\bibfnamefont {D.}~\bibnamefont
  {Hedman}}, \bibinfo {author} {\bibfnamefont {B.}~\bibnamefont {McLean}},
  \bibinfo {author} {\bibfnamefont {C.}~\bibnamefont {Bichara}}, \bibinfo
  {author} {\bibfnamefont {S.}~\bibnamefont {Maruyama}}, \bibinfo {author}
  {\bibfnamefont {J.~A.}\ \bibnamefont {Larsson}},\ and\ \bibinfo {author}
  {\bibfnamefont {F.}~\bibnamefont {Ding}},\ }\bibfield  {title} {\bibinfo
  {title} {Dynamics of growing carbon nanotube interfaces probed by machine
  learning-enabled molecular simulations},\ }\bibfield  {journal} {\bibinfo
  {journal} {arXiv preprint arXiv:2302.09542}\ }\href
  {https://doi.org/10.48550/arXiv.2302.09542} {10.48550/arXiv.2302.09542}
  (\bibinfo {year} {2023})\BibitemShut {NoStop}%
\bibitem [{\citenamefont {Zhang}\ \emph {et~al.}(2020)\citenamefont {Zhang},
  \citenamefont {Wang}, \citenamefont {Chen}, \citenamefont {Zeng},
  \citenamefont {Zhang}, \citenamefont {Wang},\ and\ \citenamefont
  {E}}]{dpgen}%
  \BibitemOpen
  \bibfield  {author} {\bibinfo {author} {\bibfnamefont {Y.}~\bibnamefont
  {Zhang}}, \bibinfo {author} {\bibfnamefont {H.}~\bibnamefont {Wang}},
  \bibinfo {author} {\bibfnamefont {W.}~\bibnamefont {Chen}}, \bibinfo {author}
  {\bibfnamefont {J.}~\bibnamefont {Zeng}}, \bibinfo {author} {\bibfnamefont
  {L.}~\bibnamefont {Zhang}}, \bibinfo {author} {\bibfnamefont
  {H.}~\bibnamefont {Wang}},\ and\ \bibinfo {author} {\bibfnamefont
  {W.}~\bibnamefont {E}},\ }\bibfield  {title} {\bibinfo {title} {{DP-GEN}: A
  concurrent learning platform for the generation of reliable deep learning
  based potential energy models},\ }\href
  {https://doi.org/10.1016/j.cpc.2020.107206} {\bibfield  {journal} {\bibinfo
  {journal} {Comput. Phys. Commun.}\ }\textbf {\bibinfo {volume} {253}},\
  \bibinfo {pages} {107206} (\bibinfo {year} {2020})}\BibitemShut {NoStop}%
\bibitem [{\citenamefont {Zhang}\ \emph {et~al.}(2018)\citenamefont {Zhang},
  \citenamefont {Han}, \citenamefont {Wang}, \citenamefont {Saidi},
  \citenamefont {Car} \emph {et~al.}}]{NEURIPS2018_e2ad76f2}%
  \BibitemOpen
  \bibfield  {author} {\bibinfo {author} {\bibfnamefont {L.}~\bibnamefont
  {Zhang}}, \bibinfo {author} {\bibfnamefont {J.}~\bibnamefont {Han}}, \bibinfo
  {author} {\bibfnamefont {H.}~\bibnamefont {Wang}}, \bibinfo {author}
  {\bibfnamefont {W.}~\bibnamefont {Saidi}}, \bibinfo {author} {\bibfnamefont
  {R.}~\bibnamefont {Car}}, \emph {et~al.},\ }\bibfield  {title} {\bibinfo
  {title} {End-to-end symmetry preserving inter-atomic potential energy model
  for finite and extended systems},\ }\href
  {https://dl.acm.org/doi/10.5555/3327345.3327356} {\bibfield  {journal}
  {\bibinfo  {journal} {Adv. Neural Inf. Process. Syst.}\ }\textbf {\bibinfo
  {volume} {31}} (\bibinfo {year} {2018})}\BibitemShut {NoStop}%
\bibitem [{\citenamefont {Wang}\ \emph {et~al.}(2018)\citenamefont {Wang},
  \citenamefont {Zhang}, \citenamefont {Han},\ and\ \citenamefont
  {E}}]{deepmd}%
  \BibitemOpen
  \bibfield  {author} {\bibinfo {author} {\bibfnamefont {H.}~\bibnamefont
  {Wang}}, \bibinfo {author} {\bibfnamefont {L.}~\bibnamefont {Zhang}},
  \bibinfo {author} {\bibfnamefont {J.}~\bibnamefont {Han}},\ and\ \bibinfo
  {author} {\bibfnamefont {W.}~\bibnamefont {E}},\ }\bibfield  {title}
  {\bibinfo {title} {{DeePMD}-kit: {A} deep learning package for many-body
  potential energy representation and molecular dynamics},\ }\href
  {https://doi.org/10.1016/j.cpc.2018.03.016} {\bibfield  {journal} {\bibinfo
  {journal} {Comput. Phys. Commun.}\ }\textbf {\bibinfo {volume} {228}},\
  \bibinfo {pages} {178} (\bibinfo {year} {2018})}\BibitemShut {NoStop}%
\bibitem [{\citenamefont {Thompson}\ \emph {et~al.}(2022)\citenamefont
  {Thompson}, \citenamefont {Aktulga}, \citenamefont {Berger}, \citenamefont
  {Bolintineanu}, \citenamefont {Brown}, \citenamefont {Crozier}, \citenamefont
  {in't Veld}, \citenamefont {Kohlmeyer}, \citenamefont {Moore}, \citenamefont
  {Nguyen} \emph {et~al.}}]{lammps}%
  \BibitemOpen
  \bibfield  {author} {\bibinfo {author} {\bibfnamefont {A.~P.}\ \bibnamefont
  {Thompson}}, \bibinfo {author} {\bibfnamefont {H.~M.}\ \bibnamefont
  {Aktulga}}, \bibinfo {author} {\bibfnamefont {R.}~\bibnamefont {Berger}},
  \bibinfo {author} {\bibfnamefont {D.~S.}\ \bibnamefont {Bolintineanu}},
  \bibinfo {author} {\bibfnamefont {W.~M.}\ \bibnamefont {Brown}}, \bibinfo
  {author} {\bibfnamefont {P.~S.}\ \bibnamefont {Crozier}}, \bibinfo {author}
  {\bibfnamefont {P.~J.}\ \bibnamefont {in't Veld}}, \bibinfo {author}
  {\bibfnamefont {A.}~\bibnamefont {Kohlmeyer}}, \bibinfo {author}
  {\bibfnamefont {S.~G.}\ \bibnamefont {Moore}}, \bibinfo {author}
  {\bibfnamefont {T.~D.}\ \bibnamefont {Nguyen}}, \emph {et~al.},\ }\bibfield
  {title} {\bibinfo {title} {{LAMMPS}: A flexible simulation tool for
  particle-based materials modeling at the atomic, meso, and continuum
  scales},\ }\href {https://doi.org/10.1016/j.cpc.2021.108171} {\bibfield
  {journal} {\bibinfo  {journal} {Comput. Phys. Commun.}\ }\textbf {\bibinfo
  {volume} {271}},\ \bibinfo {pages} {108171} (\bibinfo {year}
  {2022})}\BibitemShut {NoStop}%
\bibitem [{\citenamefont {Lee}\ \emph {et~al.}(2023)\citenamefont {Lee},
  \citenamefont {Hedman}, \citenamefont {Dong}, \citenamefont {Zhang},
  \citenamefont {Lee}, \citenamefont {Kim},\ and\ \citenamefont
  {Ding}}]{lee2023importance}%
  \BibitemOpen
  \bibfield  {author} {\bibinfo {author} {\bibfnamefont {W.}~\bibnamefont
  {Lee}}, \bibinfo {author} {\bibfnamefont {D.}~\bibnamefont {Hedman}},
  \bibinfo {author} {\bibfnamefont {J.}~\bibnamefont {Dong}}, \bibinfo {author}
  {\bibfnamefont {L.}~\bibnamefont {Zhang}}, \bibinfo {author} {\bibfnamefont
  {Z.}~\bibnamefont {Lee}}, \bibinfo {author} {\bibfnamefont {S.~Y.}\
  \bibnamefont {Kim}},\ and\ \bibinfo {author} {\bibfnamefont {F.}~\bibnamefont
  {Ding}},\ }\bibfield  {title} {\bibinfo {title} {Importance of kink energy in
  calculating the formation energy of a graphene edge},\ }\href@noop {}
  {\bibfield  {journal} {\bibinfo  {journal} {Phys. Rev. B}\ }\textbf {\bibinfo
  {volume} {107}},\ \bibinfo {pages} {245420} (\bibinfo {year}
  {2023})}\BibitemShut {NoStop}%
\bibitem [{\citenamefont {Nuismer}(1975)}]{nuismer1975}%
  \BibitemOpen
  \bibfield  {author} {\bibinfo {author} {\bibfnamefont {R.}~\bibnamefont
  {Nuismer}},\ }\bibfield  {title} {\bibinfo {title} {An energy release rate
  criterion for mixed mode fracture},\ }\href
  {https://doi.org/10.1007/BF00038891} {\bibfield  {journal} {\bibinfo
  {journal} {Int. J. Fract.}\ }\textbf {\bibinfo {volume} {11}},\ \bibinfo
  {pages} {245} (\bibinfo {year} {1975})}\BibitemShut {NoStop}%
\bibitem [{\citenamefont {Williams}(1957)}]{williams1957}%
  \BibitemOpen
  \bibfield  {author} {\bibinfo {author} {\bibfnamefont {M.~L.}\ \bibnamefont
  {Williams}},\ }\bibfield  {title} {\bibinfo {title} {On the stress
  distribution at the base of a stationary crack},\ }\href
  {https://doi.org/10.1115/1.4011454} {\bibfield  {journal} {\bibinfo
  {journal} {J. Appl. Mech.}\ }\textbf {\bibinfo {volume} {24}},\ \bibinfo
  {pages} {109} (\bibinfo {year} {1957})}\BibitemShut {NoStop}%
\bibitem [{\citenamefont {Wang}\ \emph {et~al.}(2016)\citenamefont {Wang},
  \citenamefont {Qin}, \citenamefont {Jung}, \citenamefont {Martin-Martinez},
  \citenamefont {Zhang}, \citenamefont {Buehler},\ and\ \citenamefont
  {Warner}}]{wang_2016}%
  \BibitemOpen
  \bibfield  {author} {\bibinfo {author} {\bibfnamefont {S.}~\bibnamefont
  {Wang}}, \bibinfo {author} {\bibfnamefont {Z.}~\bibnamefont {Qin}}, \bibinfo
  {author} {\bibfnamefont {G.~S.}\ \bibnamefont {Jung}}, \bibinfo {author}
  {\bibfnamefont {F.~J.}\ \bibnamefont {Martin-Martinez}}, \bibinfo {author}
  {\bibfnamefont {K.}~\bibnamefont {Zhang}}, \bibinfo {author} {\bibfnamefont
  {M.~J.}\ \bibnamefont {Buehler}},\ and\ \bibinfo {author} {\bibfnamefont
  {J.~H.}\ \bibnamefont {Warner}},\ }\bibfield  {title} {\bibinfo {title}
  {Atomically sharp crack tips in monolayer {M}o{S}$_{2}$ and their enhanced
  toughness by vacancy defects},\ }\href
  {https://doi.org/10.1021/acsnano.6b05435} {\bibfield  {journal} {\bibinfo
  {journal} {ACS Nano}\ }\textbf {\bibinfo {volume} {10}},\ \bibinfo {pages}
  {9831} (\bibinfo {year} {2016})}\BibitemShut {NoStop}%
\bibitem [{\citenamefont {Van~der Ven}\ and\ \citenamefont
  {Ceder}(2004)}]{van2004thermodynamics}%
  \BibitemOpen
  \bibfield  {author} {\bibinfo {author} {\bibfnamefont {A.}~\bibnamefont
  {Van~der Ven}}\ and\ \bibinfo {author} {\bibfnamefont {G.}~\bibnamefont
  {Ceder}},\ }\bibfield  {title} {\bibinfo {title} {The thermodynamics of
  decohesion},\ }\href {https://doi.org/10.1016/j.actamat.2003.11.007}
  {\bibfield  {journal} {\bibinfo  {journal} {Acta Mater.}\ }\textbf {\bibinfo
  {volume} {52}},\ \bibinfo {pages} {1223} (\bibinfo {year}
  {2004})}\BibitemShut {NoStop}%
\bibitem [{\citenamefont {Tada}\ \emph {et~al.}(2000)\citenamefont {Tada},
  \citenamefont {Paris},\ and\ \citenamefont {Irwin}}]{handbook}%
  \BibitemOpen
  \bibfield  {author} {\bibinfo {author} {\bibfnamefont {H.}~\bibnamefont
  {Tada}}, \bibinfo {author} {\bibfnamefont {P.~C.}\ \bibnamefont {Paris}},\
  and\ \bibinfo {author} {\bibfnamefont {G.~R.}\ \bibnamefont {Irwin}},\ }\href
  {https://doi.org/10.1115/1.801535} {\emph {\bibinfo {title} {{The Stress
  Analysis of Cracks Handbook, Third Edition}}}}\ (\bibinfo  {publisher} {ASME
  Press},\ \bibinfo {year} {2000})\BibitemShut {NoStop}%
\bibitem [{\citenamefont {Ayatollahi}\ and\ \citenamefont
  {Nejati}(2011)}]{over-determined}%
  \BibitemOpen
  \bibfield  {author} {\bibinfo {author} {\bibfnamefont {M.}~\bibnamefont
  {Ayatollahi}}\ and\ \bibinfo {author} {\bibfnamefont {M.}~\bibnamefont
  {Nejati}},\ }\bibfield  {title} {\bibinfo {title} {An over-deterministic
  method for calculation of coefficients of crack tip asymptotic field from
  finite element analysis},\ }\href
  {https://doi.org/10.1111/j.1460-2695.2010.01504.x} {\bibfield  {journal}
  {\bibinfo  {journal} {Fatigue $\&$ Fract. Eng. Mater. Struct.}\ }\textbf
  {\bibinfo {volume} {34}},\ \bibinfo {pages} {159} (\bibinfo {year}
  {2011})}\BibitemShut {NoStop}%
\bibitem [{\citenamefont {Wilson}\ \emph {et~al.}(2019)\citenamefont {Wilson},
  \citenamefont {Grutzik},\ and\ \citenamefont {Chandross}}]{wilson_2019}%
  \BibitemOpen
  \bibfield  {author} {\bibinfo {author} {\bibfnamefont {M.~A.}\ \bibnamefont
  {Wilson}}, \bibinfo {author} {\bibfnamefont {S.~J.}\ \bibnamefont
  {Grutzik}},\ and\ \bibinfo {author} {\bibfnamefont {M.}~\bibnamefont
  {Chandross}},\ }\bibfield  {title} {\bibinfo {title} {Continuum stress
  intensity factors from atomistic fracture simulations},\ }\href
  {https://doi.org/10.1016/j.cma.2019.05.050} {\bibfield  {journal} {\bibinfo
  {journal} {Comput. Methods Appl. Mech. Engrg.}\ }\textbf {\bibinfo {volume}
  {354}},\ \bibinfo {pages} {732} (\bibinfo {year} {2019})}\BibitemShut
  {NoStop}%
\bibitem [{\citenamefont {Soler}\ \emph {et~al.}(2002)\citenamefont {Soler},
  \citenamefont {Artacho}, \citenamefont {Gale}, \citenamefont {Garc{\'\i}a},
  \citenamefont {Junquera}, \citenamefont {Ordej{\'o}n},\ and\ \citenamefont
  {S{\'a}nchez-Portal}}]{siesta}%
  \BibitemOpen
  \bibfield  {author} {\bibinfo {author} {\bibfnamefont {J.~M.}\ \bibnamefont
  {Soler}}, \bibinfo {author} {\bibfnamefont {E.}~\bibnamefont {Artacho}},
  \bibinfo {author} {\bibfnamefont {J.~D.}\ \bibnamefont {Gale}}, \bibinfo
  {author} {\bibfnamefont {A.}~\bibnamefont {Garc{\'\i}a}}, \bibinfo {author}
  {\bibfnamefont {J.}~\bibnamefont {Junquera}}, \bibinfo {author}
  {\bibfnamefont {P.}~\bibnamefont {Ordej{\'o}n}},\ and\ \bibinfo {author}
  {\bibfnamefont {D.}~\bibnamefont {S{\'a}nchez-Portal}},\ }\bibfield  {title}
  {\bibinfo {title} {The {SIESTA} method for ab initio order-{$N$} materials
  simulation},\ }\href {https://doi.org/10.1088/0953-8984/14/11/302} {\bibfield
   {journal} {\bibinfo  {journal} {J. Phys. Condens. Matter.}\ }\textbf
  {\bibinfo {volume} {14}},\ \bibinfo {pages} {2745} (\bibinfo {year}
  {2002})}\BibitemShut {NoStop}%
\bibitem [{\citenamefont {Perdew}\ \emph {et~al.}(1996)\citenamefont {Perdew},
  \citenamefont {Burke},\ and\ \citenamefont {Ernzerhof}}]{PBE}%
  \BibitemOpen
  \bibfield  {author} {\bibinfo {author} {\bibfnamefont {J.~P.}\ \bibnamefont
  {Perdew}}, \bibinfo {author} {\bibfnamefont {K.}~\bibnamefont {Burke}},\ and\
  \bibinfo {author} {\bibfnamefont {M.}~\bibnamefont {Ernzerhof}},\ }\bibfield
  {title} {\bibinfo {title} {Generalized gradient approximation made simple},\
  }\href {https://doi.org/10.1103/PhysRevLett.77.3865} {\bibfield  {journal}
  {\bibinfo  {journal} {Phys. Rev. Lett.}\ }\textbf {\bibinfo {volume} {77}},\
  \bibinfo {pages} {3865} (\bibinfo {year} {1996})}\BibitemShut {NoStop}%
\bibitem [{\citenamefont {Troullier}\ and\ \citenamefont
  {Martins}(1991)}]{troullier1991}%
  \BibitemOpen
  \bibfield  {author} {\bibinfo {author} {\bibfnamefont {N.}~\bibnamefont
  {Troullier}}\ and\ \bibinfo {author} {\bibfnamefont {J.~L.}\ \bibnamefont
  {Martins}},\ }\bibfield  {title} {\bibinfo {title} {Efficient
  pseudopotentials for plane-wave calculations},\ }\href
  {https://doi.org/10.1103/PhysRevB.43.1993} {\bibfield  {journal} {\bibinfo
  {journal} {Phys. Rev. B}\ }\textbf {\bibinfo {volume} {43}},\ \bibinfo
  {pages} {1993} (\bibinfo {year} {1991})}\BibitemShut {NoStop}%
\bibitem [{\citenamefont {Kresse}\ and\ \citenamefont
  {Furthm{\"u}ller}(1996)}]{VASP}%
  \BibitemOpen
  \bibfield  {author} {\bibinfo {author} {\bibfnamefont {G.}~\bibnamefont
  {Kresse}}\ and\ \bibinfo {author} {\bibfnamefont {J.}~\bibnamefont
  {Furthm{\"u}ller}},\ }\bibfield  {title} {\bibinfo {title} {Efficient
  iterative schemes for ab initio total-energy calculations using a plane-wave
  basis set},\ }\href {https://doi.org/10.1103/PhysRevB.54.11169} {\bibfield
  {journal} {\bibinfo  {journal} {Phys. Rev. B}\ }\textbf {\bibinfo {volume}
  {54}},\ \bibinfo {pages} {11169} (\bibinfo {year} {1996})}\BibitemShut
  {NoStop}%
\bibitem [{\citenamefont {Bl{\"o}chl}(1994)}]{PAW}%
  \BibitemOpen
  \bibfield  {author} {\bibinfo {author} {\bibfnamefont {P.~E.}\ \bibnamefont
  {Bl{\"o}chl}},\ }\bibfield  {title} {\bibinfo {title} {Projector
  augmented-wave method},\ }\href {https://doi.org/10.1103/PhysRevB.50.17953}
  {\bibfield  {journal} {\bibinfo  {journal} {Phys. Rev. B}\ }\textbf {\bibinfo
  {volume} {50}},\ \bibinfo {pages} {17953} (\bibinfo {year}
  {1994})}\BibitemShut {NoStop}%
\bibitem [{\citenamefont {Addou}\ \emph {et~al.}(2018)\citenamefont {Addou},
  \citenamefont {Smyth}, \citenamefont {Noh}, \citenamefont {Lin},
  \citenamefont {Pan}, \citenamefont {Eichfeld}, \citenamefont {F{\"o}lsch},
  \citenamefont {Robinson}, \citenamefont {Cho}, \citenamefont {Feenstra} \emph
  {et~al.}}]{addou_2018}%
  \BibitemOpen
  \bibfield  {author} {\bibinfo {author} {\bibfnamefont {R.}~\bibnamefont
  {Addou}}, \bibinfo {author} {\bibfnamefont {C.~M.}\ \bibnamefont {Smyth}},
  \bibinfo {author} {\bibfnamefont {J.-Y.}\ \bibnamefont {Noh}}, \bibinfo
  {author} {\bibfnamefont {Y.-C.}\ \bibnamefont {Lin}}, \bibinfo {author}
  {\bibfnamefont {Y.}~\bibnamefont {Pan}}, \bibinfo {author} {\bibfnamefont
  {S.~M.}\ \bibnamefont {Eichfeld}}, \bibinfo {author} {\bibfnamefont
  {S.}~\bibnamefont {F{\"o}lsch}}, \bibinfo {author} {\bibfnamefont {J.~A.}\
  \bibnamefont {Robinson}}, \bibinfo {author} {\bibfnamefont {K.}~\bibnamefont
  {Cho}}, \bibinfo {author} {\bibfnamefont {R.~M.}\ \bibnamefont {Feenstra}},
  \emph {et~al.},\ }\bibfield  {title} {\bibinfo {title} {One dimensional
  metallic edges in atomically thin {WS}e$_{2}$ induced by air exposure},\
  }\href {https://doi.org/10.1088/2053-1583/aab0cd} {\bibfield  {journal}
  {\bibinfo  {journal} {2D Mater.}\ }\textbf {\bibinfo {volume} {5}},\ \bibinfo
  {pages} {025017} (\bibinfo {year} {2018})}\BibitemShut {NoStop}%
\bibitem [{\citenamefont {Xiao}\ \emph {et~al.}(2016)\citenamefont {Xiao},
  \citenamefont {Yu},\ and\ \citenamefont {Gao}}]{xiao_2016}%
  \BibitemOpen
  \bibfield  {author} {\bibinfo {author} {\bibfnamefont {S.-L.}\ \bibnamefont
  {Xiao}}, \bibinfo {author} {\bibfnamefont {W.-Z.}\ \bibnamefont {Yu}},\ and\
  \bibinfo {author} {\bibfnamefont {S.-P.}\ \bibnamefont {Gao}},\ }\bibfield
  {title} {\bibinfo {title} {Edge preference and band gap characters of
  {M}o{S}$_{2}$ and {WS}$_{2}$ nanoribbons},\ }\href
  {https://doi.org/10.1016/j.susc.2016.06.011} {\bibfield  {journal} {\bibinfo
  {journal} {Surf. Sci.}\ }\textbf {\bibinfo {volume} {653}},\ \bibinfo {pages}
  {107} (\bibinfo {year} {2016})}\BibitemShut {NoStop}%
\bibitem [{\citenamefont {Lee}\ \emph {et~al.}(2017)\citenamefont {Lee},
  \citenamefont {Yoon}, \citenamefont {Scullion}, \citenamefont {Jang},
  \citenamefont {Santos}, \citenamefont {Jeong},\ and\ \citenamefont
  {Kim}}]{lee_2017}%
  \BibitemOpen
  \bibfield  {author} {\bibinfo {author} {\bibfnamefont {Y.}~\bibnamefont
  {Lee}}, \bibinfo {author} {\bibfnamefont {J.-Y.}\ \bibnamefont {Yoon}},
  \bibinfo {author} {\bibfnamefont {D.}~\bibnamefont {Scullion}}, \bibinfo
  {author} {\bibfnamefont {J.}~\bibnamefont {Jang}}, \bibinfo {author}
  {\bibfnamefont {E.~J.}\ \bibnamefont {Santos}}, \bibinfo {author}
  {\bibfnamefont {H.~Y.}\ \bibnamefont {Jeong}},\ and\ \bibinfo {author}
  {\bibfnamefont {K.}~\bibnamefont {Kim}},\ }\bibfield  {title} {\bibinfo
  {title} {Atomic-scale imaging of few-layer black phosphorus and its
  reconstructed edge},\ }\href {https://doi.org/10.1088/1361-6463/aa5583}
  {\bibfield  {journal} {\bibinfo  {journal} {J. Phys. D: Appl. Phys.}\
  }\textbf {\bibinfo {volume} {50}},\ \bibinfo {pages} {084003} (\bibinfo
  {year} {2017})}\BibitemShut {NoStop}%
\bibitem [{\citenamefont {Aghaei}\ \emph {et~al.}(2016)\citenamefont {Aghaei},
  \citenamefont {Monshi},\ and\ \citenamefont {Calizo}}]{aghaei_2016}%
  \BibitemOpen
  \bibfield  {author} {\bibinfo {author} {\bibfnamefont {S.}~\bibnamefont
  {Aghaei}}, \bibinfo {author} {\bibfnamefont {M.}~\bibnamefont {Monshi}},\
  and\ \bibinfo {author} {\bibfnamefont {I.}~\bibnamefont {Calizo}},\
  }\bibfield  {title} {\bibinfo {title} {A theoretical study of gas adsorption
  on silicene nanoribbons and its application in a highly sensitive molecule
  sensor},\ }\href {https://doi.org/10.1039/c6ra21293j} {\bibfield  {journal}
  {\bibinfo  {journal} {RSC Adv.}\ }\textbf {\bibinfo {volume} {6}},\ \bibinfo
  {pages} {94417} (\bibinfo {year} {2016})}\BibitemShut {NoStop}%
\bibitem [{\citenamefont {Zhou}\ \emph {et~al.}(2013)\citenamefont {Zhou},
  \citenamefont {Zou}, \citenamefont {Najmaei}, \citenamefont {Liu},
  \citenamefont {Shi}, \citenamefont {Kong}, \citenamefont {Lou}, \citenamefont
  {Ajayan}, \citenamefont {Yakobson},\ and\ \citenamefont
  {Idrobo}}]{zhou_2013}%
  \BibitemOpen
  \bibfield  {author} {\bibinfo {author} {\bibfnamefont {W.}~\bibnamefont
  {Zhou}}, \bibinfo {author} {\bibfnamefont {X.}~\bibnamefont {Zou}}, \bibinfo
  {author} {\bibfnamefont {S.}~\bibnamefont {Najmaei}}, \bibinfo {author}
  {\bibfnamefont {Z.}~\bibnamefont {Liu}}, \bibinfo {author} {\bibfnamefont
  {Y.}~\bibnamefont {Shi}}, \bibinfo {author} {\bibfnamefont {J.}~\bibnamefont
  {Kong}}, \bibinfo {author} {\bibfnamefont {J.}~\bibnamefont {Lou}}, \bibinfo
  {author} {\bibfnamefont {P.~M.}\ \bibnamefont {Ajayan}}, \bibinfo {author}
  {\bibfnamefont {B.~I.}\ \bibnamefont {Yakobson}},\ and\ \bibinfo {author}
  {\bibfnamefont {J.-C.}\ \bibnamefont {Idrobo}},\ }\bibfield  {title}
  {\bibinfo {title} {Intrinsic structural defects in monolayer molybdenum
  disulfide},\ }\href {https://doi.org/10.1021/nl4007479} {\bibfield  {journal}
  {\bibinfo  {journal} {Nano Lett.}\ }\textbf {\bibinfo {volume} {13}},\
  \bibinfo {pages} {2615} (\bibinfo {year} {2013})}\BibitemShut {NoStop}%
\bibitem [{\citenamefont {Yang}\ \emph {et~al.}(2017)\citenamefont {Yang},
  \citenamefont {Li}, \citenamefont {Wen}, \citenamefont {Hacopian},
  \citenamefont {Chen}, \citenamefont {Gong}, \citenamefont {Zhang},
  \citenamefont {Li}, \citenamefont {Zhou}, \citenamefont {Ajayan} \emph
  {et~al.}}]{yang_2017}%
  \BibitemOpen
  \bibfield  {author} {\bibinfo {author} {\bibfnamefont {Y.}~\bibnamefont
  {Yang}}, \bibinfo {author} {\bibfnamefont {X.}~\bibnamefont {Li}}, \bibinfo
  {author} {\bibfnamefont {M.}~\bibnamefont {Wen}}, \bibinfo {author}
  {\bibfnamefont {E.}~\bibnamefont {Hacopian}}, \bibinfo {author}
  {\bibfnamefont {W.}~\bibnamefont {Chen}}, \bibinfo {author} {\bibfnamefont
  {Y.}~\bibnamefont {Gong}}, \bibinfo {author} {\bibfnamefont {J.}~\bibnamefont
  {Zhang}}, \bibinfo {author} {\bibfnamefont {B.}~\bibnamefont {Li}}, \bibinfo
  {author} {\bibfnamefont {W.}~\bibnamefont {Zhou}}, \bibinfo {author}
  {\bibfnamefont {P.~M.}\ \bibnamefont {Ajayan}}, \emph {et~al.},\ }\bibfield
  {title} {\bibinfo {title} {Brittle fracture of {2D} {M}o{S}e$_{2}$},\ }\href
  {https://doi.org/10.1002/adma.201604201} {\bibfield  {journal} {\bibinfo
  {journal} {Adv. Mater.}\ }\textbf {\bibinfo {volume} {29}},\ \bibinfo {pages}
  {1604201} (\bibinfo {year} {2017})}\BibitemShut {NoStop}%
\bibitem [{\citenamefont {Hoogenboom}(2021)}]{AFM_2021}%
  \BibitemOpen
  \bibfield  {author} {\bibinfo {author} {\bibfnamefont {B.~W.}\ \bibnamefont
  {Hoogenboom}},\ }\bibfield  {title} {\bibinfo {title} {Stretching the
  resolution limit of atomic force microscopy},\ }\href
  {https://doi.org/10.1038/s41594-021-00638-x} {\bibfield  {journal} {\bibinfo
  {journal} {Nat. Struct. Mol. Biol.}\ }\textbf {\bibinfo {volume} {28}},\
  \bibinfo {pages} {629} (\bibinfo {year} {2021})}\BibitemShut {NoStop}%
\bibitem [{\citenamefont {Tsuno}(1993)}]{TEM_1993}%
  \BibitemOpen
  \bibfield  {author} {\bibinfo {author} {\bibfnamefont {K.}~\bibnamefont
  {Tsuno}},\ }\bibfield  {title} {\bibinfo {title} {Resolution limit of a
  transmission electron microscope with an uncorrected conventional magnetic
  objective lens},\ }\href {https://doi.org/10.1016/0304-3991(93)90193-2}
  {\bibfield  {journal} {\bibinfo  {journal} {Ultramicroscopy}\ }\textbf
  {\bibinfo {volume} {50}},\ \bibinfo {pages} {245} (\bibinfo {year}
  {1993})}\BibitemShut {NoStop}%
\bibitem [{\citenamefont {Shi}\ \emph {et~al.}(2023)\citenamefont {Shi},
  \citenamefont {Feng},\ and\ \citenamefont {Xu}}]{dataset}%
  \BibitemOpen
  \bibfield  {author} {\bibinfo {author} {\bibfnamefont {P.}~\bibnamefont
  {Shi}}, \bibinfo {author} {\bibfnamefont {S.}~\bibnamefont {Feng}},\ and\
  \bibinfo {author} {\bibfnamefont {Z.}~\bibnamefont {Xu}},\ }\bibfield
  {title} {\bibinfo {title} {Non-equilibrium nature of fracture determines the
  crack path},\ }\bibfield  {journal} {\bibinfo  {journal} {Materials Cloud
  Archive}\ }\textbf {\bibinfo {volume} {2023.X}},\ \href
  {https://doi.org/10.24435/materialscloud:rd-0e}
  {10.24435/materialscloud:rd-0e} (\bibinfo {year} {2023})\BibitemShut
  {NoStop}%
\end{thebibliography}%

\end{document}